# Magnetospheric Multiscale Dayside Reconnection Electron Diffusion Region Events

J. M. Webster[1], J. L. Burch[2], P. H. Reiff[1], D. B. Graham[3], R. B. Torbert[4], R. E. Ergun[5], A. G. Daou[1], S. Y. Sazykin[1], A. Marshall[1], R.C. Allen[6], L.-J. Chen[7], S. Wang[7], T. D. Phan[8], K. J. Genestreti[9], B. L. Giles[10], T. E. Moore[10], S. A. Fuselier[2,11], G. Cozzani[12], C. T. Russell[13], S. Eriksson[5], A. C. Rager[10], J. M. Broll[11,2], K. Goodrich[5], F. Wilder[5]

[1]Rice University, Houston, TX, USA
[2]Southwest Research Institute, San Antonio, TX, USA
[3]Swedish Institute of Space Physics, Uppsala, Sweden
[4]University of New Hampshire, Durham, NH, USA
[5]University of Colorado, LASP, Boulder, CO, USA
[6]Johns Hopkins University, APL, Laurel, MD, USA
[7]University of Maryland, College Park, MD, USA
[8]University of California, Berkeley, CA, USA
[9]Austrian Academy of Sciences, Graz, Austria
[10]NASA, Goddard Space Flight Center, Greenbelt, MD
[11]University of Texas at San Antonio, San Antonio, TX, USA
[12]Pierre and Marie Curie University, Paris, France
[13]University of California at Los Angeles, Los Angeles, CA, USA

Corresponding author: James Webster (james.m.webster@rice.edu)

## Abstract

We have used the high-resolution data of the Magnetospheric Multiscale (MMS) mission's dayside phase to identify twenty-one previously unreported encounters with the electron diffusion region (EDR), as evidenced by electron agyrotropy, ion jet reversals, and **j** * **E'** > 0. Three of the new EDR encounters, which occurred within a one-minute-long interval on November 23$^{rd}$, 2016 are analyzed in detail. These events, which resulted from a relatively low and oscillating magnetopause velocity, contained large electric fields (several tens to hundreds of mV/m), crescent-shaped electron velocity phase space densities, large currents ($\geq 2$ µA/m$^2$), and Ohmic heating of the plasma (~10 nW/m$^3$). Because of the slow in-and-out motion of the magnetopause, two of these events show the unprecedented mixture of perpendicular and parallel crescents, indicating the first breaking and reconnecting of solar wind and magnetospheric field lines. An extended list of thirty-two EDR or near-EDR events is also included, and demonstrates a wide variety of observed plasma behavior inside and surrounding the reconnection site.

**Key Points:**

**(i) Multiple encounters of the EDR shown co-existence of perpendicular and parallel electron crescent distributions.**

**(ii) Ion jet reversals coincide with reversals in direction of electron crescent distributions.**

**(iii) Thirty-two EDR or near-EDR encounters exhibit significant differences in plasma properties.**

**1. Introduction**

The MMS instrumentation suite utilizes naturally-occurring Sun-Earth interactions as a laboratory in which to study magnetic reconnection. The pursuit of in-situ measurements of electron motion around and inside the EDR drove much of the mission's motivation and design [*Burch et al.*, 2016a]. From one of the first EDR encounters, *Burch et al.* [2016] reported agyrotropic, crescent-shaped electron distribution functions in the plane normal to the magnetic field vector. Such crescents are a direct observational indication of an EDR [*Hesse et al.*, 2014], and occur primarily in the region near the flow stagnation point. *Burch et al.* [2016b] further reported the evolution of perpendicular crescents into parallel crescents as the magnetic field lines from the solar wind and the magnetosphere reconnected, forming open field lines in the region between the stagnation point and the X-line.

Crescent-shaped electron PSDs and associated EDR physics have previously been reported in twelve MMS encounters, including *Chen et al.* [2016], *Norgren et al.* [2016], *Khotyaintsev et al.* [2016], and others. All twelve were recently reviewed by *Fuselier et al.* [2017], although we do not include *Eriksson et al.'s* [2016] Kelvin-Helmholtz EDR event in this analysis. To continue building upon the established collection, additional EDRs and EDR candidates exhibiting crescent-like electron PSD shapes were identified in the MMS dataset. In this study, we first searched for dayside events where the particle and field data indicate a magnetopause crossing with large $j_y$, small $|\mathbf{B}|$, electron heating, and ion jet reversals, and examined the high-resolution electron distributions for crescent-shaped enhancements. Fewer than one crossing in fifteen exhibited these crescents. We acknowledge that imposing the constraint of a small B-field strength may bias the selections towards small guide field reconnection (small $B_M$). Checking for this bias will require further analysis to obtain appropriate boundary-normal coordinates for each new event.

A fortunate series of direct EDR encounters on November 23rd, 2016 yielded three new events within a span of ~1 minute, for which we provide an introductory analysis here. During or immediately bordering these three events, the MMS data exhibited several established EDR signatures:

1) Electron heating parallel to the magnetic field, similar to previous THEMIS [*Tang et al.*, 2013] and Cluster [*Hwang et al.*, 2013] measurements taken near EDRs.

2) Sustained $+E_N$, simultaneous with i) a higher-energy, minority population of electrons traveling perpendicular to **B**, ii) agyrotropy (*Swisdak et al.*, [2016]), and iii) crescent distributions.

3) Observations of Ohmic dissipation via electromagnetic fields in the form of **j** * **E'**, where **E'** = **E** + **v**$_e$ x **B**, all separately measured quantities.

A moderate guide field ($B_M/B_L$ ~1/2) was seen during the first encounter, which took place at the magnetic *X* line and surrounding magnetosheath. Notable gradients in the magnetic field and temperature were also present. The next two EDR events occurred on the magnetospheric side of the region, and also show $B_M/B_L$ ~1/2.

During the first of the two magnetosphere-side EDR encounters on November 23$^{rd}$, 2016, especially large amplitude electrostatic waves were recorded, with peak values of ~100 mV/m. *Ergun et al.* [2016a] reported similar observations. The E-field oscillations are oriented primarily along **B**, and may be indicative of electron bunches and/or holes propagating along the separatrix [*Drake et al.*, 2003]. Here, some electrostatic waves show possible evidence of electron sheets trapped within them, akin to those seen by *Kellogg et al.* [2010], in which E-field waveforms of electrostatic whistlers contain a secondary perturbation from sheets of cold electrons undergoing transport.

An important new result in two of the three successive events is the continuoius co-existence of perpendicular and parallel crescents on the same field line and the reversal of the direction of parallel crescents that is correlated with an ion jet reversal, suggesting the motion of the spacecraft through the heart of an EDR.

In an effort to spur future studies, we also provide a quick preview of another additional eighteen EDR candidate events, listed here for the first time, bringing our total number of candidate events to thirty-two. We plot one selected agyrotropic electron crescent-like PSD from each. Our selections then undergo a standardized set of computations designed to serve as a preliminary meta-analysis, presented in a table. One example of a simple correlative comparison is included.

## 2. Observations

Each 3D electron phase space distribution (PSD) shown in this study is built from the total electron flux accrued over one 30 ms interval [*Pollock et al.*, 2016]. Every PSD plot shows the volume confined inside of a 20° half-angle cone (focal point at v=0) cross-sectional area revolved 360° around the axis pointing in/out of the page. Time-aliasing effects create jagged edges in octagonal spoke-like patterns, and indicate a distribution changing on timescales faster than the 30 ms acquisition window. The listed times above all PSDs represent the middle of the acquisition window, i.e. 15 ms after each new 30 ms window begins. All PSDs are shown in the rest frame of the spacecraft, with dotted lines designating the computed electron bulk velocity components projected onto the viewing planes. Color contour scaling of the PSD plots is kept constant throughout this publication. The $v_{\perp 1}$ direction for all PSD figures is defined by the average (taken over the 30 ms acquisition window) electron bulk velocity component perpendicular to the local magnetic field direction. To assist our visual PSD comparisons with a numerical indicator, we frequently list *Swisdak et al.*'s [2016] agyrotropy index, which is defined as:

$$Q = \frac{P_{12}^2 + P_{13}^2 + P_{23}^2}{P_\perp^2 + 2 P_\perp P_\parallel},$$

where the pressure tensor is:

$$\mathbb{P} = \begin{pmatrix} P_\parallel & P_{12} & P_{13} \\ P_{12} & P_\perp & P_{23} \\ P_{13} & P_{23} & P_\perp \end{pmatrix}.$$

The scaling of $\sqrt{Q_e}$ (subscript "e" denoting electrons) ranges from 0 (no agyrotropy) to 1 (total agyrotropy), and should assume comparatively large values near the EDR. Defining the perpendicular and parallel directions requires MMS's magnetic field data [*Le Contel et al.*, 2014; *Russell et al.*, 2016], and we also invoke electric field measurements [*Lindqvist et al.*, 2016; *Ergun et al.*, 2016b; *Torbert et al.*, 2016] for many analyses.

## 3. The November 23rd, 2016 EDR Events

On November 23$^{rd}$, 2016, the MMS spacecraft trajectory intersected the electron diffusion region several times. Widely-accepted features of EDRs were seen during at least three instances, including thin current sheets [*Drake et al.*, 1994], small |**B**|, significant Ohmic dissipation (**j** * **E**'), large √$Q_e$, notable wave activity, and crescent-shaped electron velocity distributions.

### 3.1) Conditions

We first review the large-scale conditions, location and timing of the event, and the spacecraft constellation configuration, using **Figure 1**. On November 23$^{rd}$, 2016, at approximately 10 minutes before 08:00 UT, Figure 1a shows MMS near the nominal magnetopause (thin line in beige band) while traveling outbound, several Earth radii duskward of the subsolar point in the Geocentric Solar Ecliptic (GSE) x-y plane. Figure 1b depicts a steady dynamic pressure supplied by the solar wind, and a significant, sustained southward interplanetary magnetic field (IMF) component ($B_z$ ~ -2 nT), given in Geocentric Solar Magnetospheric (GSM) coordinates. The steadiness of the flow pressure and magnetic field acted to confine the EDR to within a relatively small volume.

For the November 23$^{rd}$, 2016 analysis, we transform the vector data into a boundary-normal coordinate system. Our coordinate transformation was obtained using the "Minimization of Faraday Residue" (MFR) method [*Khrabrov & Sonnerup*, 1998]. The transform, given in base GSM coordinates, is L = [.317, .391, .864], M = [.264, -.911, .316], N = [.917, .136, -.373], and was found by averaging the results of the MFR calculation for the four individual spacecraft, performed over the interval from 07:49:40 UT to 07:50:15 UT. Using this transform, spacecraft relative positions are then rotated into boundary-normal coordinates in Figure 1c, with the location of the centroid of the constellation defining the plot's origin. For the thin current sheet structures near the EDR, relative spacecraft positions along the N-axis should play an especially important role in the observations. Note that MMS1 and MMS4 differed by less than 1 km in their N-coordinates, while MMS2 sampled conditions ~5 km Earthward (-N direction), with MMS3 located in between. The use of this coordinate system throughout Section 3 will aid in comparisons between the three EDR events.

### 3.2) Overview

A 1.5-minute overview plot of the event is shown in **Figure 2**, displaying the data from MMS3, as it represents the location nearest the centroid in the N-direction. Over the span of the plot, we see that the MMS constellation began in the magnetosheath (negative $B_L$, large $n_e$), passed into the magnetosphere near 07:49:35 UT, and then eventually returned back to the sheath by ~07:50:40 UT. Electric and magnetic field waves were strong, particularly near the lower hybrid frequency ($f_{lh}$), but also extending above the electron cyclotron frequency ($f_{ce}$) for the E-field. The E-field waves showed the largest amplitudes and frequencies for durations of time that the spacecraft were inside the magnetosphere and nearest an energetic EDR, at approximately 07:49:52 UT. Rapid-onset ion jet reversals [*Petrinec et al.*, 2016], agyrotropy in both ions and electrons, heated populations, and large **j** and **j** * **E**' all occurred inside of our three designated EDR events. Although many more interesting signatures exist throughout the 90 s plot, we will concentrate on observations confined within these three intervals.

### 3.3) Event "1"

The first encounter of the sequence occurred at 07:49:33 UT, when the EDR location moved southward towards the spacecraft, inferred by the ion jet of large -$v_{iL}$ values sharply tapering towards zero in panel (v) of Figure 2. MMS passed through the in-plane magnetic null point, and then moved into the magnetosheath immediately bordering an active EDR, as seen in **Figure 3**. Figure 3a plots three seconds of MMS2 data. Ohmic dissipation measurements logged several intervals of prolonged non-zero values surrounding the middle region of relatively low-density ($n_e$ ~9 cm$^{-3}$) magnetosheath plasma ($B_L$ ~ -20 nT), with significant **j** simultaneously. This middle region was host to an electron population that had undergone notable cooling in its perpendicular velocity components. The cooling results from inflowing sheath electrons conserving their first adiabatic invariant as they approach the *X* line near the border of the EDR, which preferentially depletes the higher-energy perpendicular (to **B**) particle populations of magnetosheath electrons approaching the magnetopause boundary: Electrons with comparatively larger gyroradii will intersect the boundary further away than electrons of lower $v_{e\perp}$ that share the same guiding center. The higher perpendicular temperature electrons with orbits intersecting the boundary layer are then lost via diamagnetic drift. Between 07:49:32.9 and :33.0 UT, MMS2 passed through the *X* line, just before seeing an easily distinguishable crescent-shaped electron PSD enhancement. The comparatively energetic crescent population is especially obvious here due to a contrast against the magnetosheath's colder perpendicular electrons we have remarked on previously. The timing of the MMS2 PSD shown in panel (i) of Figure 3b is indicated by the vertical, dashed line in Figure 3a. The agyrotropy listed here for MMS2 is the largest known $\sqrt{Q_e}$ recorded during the MMS mission so far. This maximum agyrotropy occurred one 30 ms window after the smallest $n_e$ measured during the 3 s interval. Figure 3b shows four electron PSDs, one from every spacecraft. The MMS1, 3, and 4 electron PSDs are chosen to be the nearest available to MMS2's, and the distributions show a near-total absence of the energetic population seen by MMS2 at this particular instant, apart from small hints in MMS3. The plots imply strong similarities in the electron populations that were traveling along the magnetic field axis. Elongation of the PSDs along **B** indicates the dominant $T_{e\parallel}$ component. We can infer that MMS2, lying furthest Earthward, was observing a localized energetic electron population at and immediately surrounding the *X* line. These energetic electrons traveling perpendicular to the local B-field did not extend outwards to the other spacecraft located only several km further sunward. Wave activity elevated from background levels was present in both **E** & **B**, visible in the spectrogram, but a pulsing at ~10 Hz (below the lower cutoff of the spectrograms) requires closer consideration. We observe that the oscillations were primarily in **B** and $T_e$, with the two temperature components closely anti-correlated. The 10 Hz waves were seen with greatest amplitudes surrounding the onset of the most negative values of $j_M$. Our November 23$^{rd}$, 2016 event is another instance of the EDR oscillation mode proposed by *Ergun et al.* [2017], which featured an EDR encounter on December 14$^{th}$, 2015. We note that the Dec 14 event oscillations were observed as the magnetopause expanded sunward, whereas this new occurrence we report on here featured a magnetopause boundary traveling earthward. The Dec 14 event was also analyzed by *Chen et al.* [2017] and *Graham et al.* [2017b].

The plot in **Figure 4** provides a multi-spacecraft picture of a 4-second interval, beginning 2 s before MMS2 saw the in-plane magnetic null. The data for the four spacecraft are color-coded. We notice a similarity of spatial gradients in multiple panels. Generally, MMS1 and MMS4 observations were the most identical, with MMS2's results comparatively further from the pair's than MMS3's, corresponding neatly to their relative separations in the N-direction. The components of the magnetic field exhibiting gradients were typically limited to $B_L$ and $B_M$.

MMS observed a density gradient pointing Earthward (larger $n_e$ for MMS2 vs. MMS1 or MMS4), but at ion length scales, the plasma density gradient across the reconnection site is known to point sunward, thus we conclude again that MMS sampled an electron-scale structure. Our previous interpretation of MMS2 residing furthest inside the EDR is also supported by the comparatively high $T_{e\perp}$ and $\sqrt{Q_e}$ values seen by MMS2 between 07:39:32.7 – :33.5 UT, although MMS1 and MMS4 observed higher **j** * **E'** values during this same interval, and with very close agreement. All four spacecraft observed EDR oscillations at ~10 Hz (best seen in $T_e$), but only MMS2 observed a large 10 Hz variation in $B_L$, setting an upper scale size of those waves to the separation (~5 km). The waves were somewhat asynchronous between spacecraft, but it is unclear how much of the asynchronicity is due to differences in particle sampling cadences of the instruments. Figure 4 also allows a rough estimate of the two boundary crossing velocities using features in **B**, assuming a static B-field topology over the interval. We use the relative time delays between $B_M \sim$ -10 nT for MMS1 and MMS4 vs. MMS2, at :32.7 vs. :32.85, respectively, and we use a 5 km separation in the N-direction, yielding an Earthward magnetopause velocity of approximately 30 km/s as $B_L$ reversed to negative values. Using the same technique with the most negative values of $B_L$ for each spacecraft in the plot (near :33.85 vs. :33.90 for MMS2 vs. MMS1, respectively), we estimate that the magnetopause moved sunward much faster, ~90 km/s, as the spacecraft re-entered the magnetosphere only 1 s after visiting the sheath. The steep temporal gradient as $B_L$ reversed again to positive values also indicates a faster sunward magnetopause velocity for the second boundary transit. Opportunities remain for many more analyses of this interval.

**3.4) Event "2"**

Following Event 1, MMS resided in the magnetopause south of the reconnection system another 20 s ($-v_{iL}$ values). MMS then made a direct approach to the EDR from the magnetospheric side, indicated by another very abrupt cessation of the $-v_{iL}$ ion jet at ~07:49:51 UT (see Figure 2). **Figure 5** shows Event 2, using 1.5 seconds of MMS1 data. MMS1 observed large $T_{e\parallel}/T_{e\perp}$ (~3) for 0.5 s at the beginning of Figure 5a. The high temperature electrons flowing along the magnetic field produced a current, as reflected in the large $+j_L$, signifying electron-scale behavior. The positive polarity of $j_L$ equates to electrons streaming in the -L direction, and we also see a large out-of-plane current ($-j_M$), indicating electrons streaming dawnward, forming the electron population of the reconnection current sheet. Figure 5a also shows especially large sustained **j** * **E**' (>2 nW/m$^{-3}$ for ~.5 s) and a high baseline of agyrotropy. MMS1 resided at or very near the electron flow stagnation plane, indicated by the steep rise in density, from 6 to 12 cm$^{-3}$, in less than 1.5 s. Over the sequence of the electron PSD plots shown in Figure 5b, MMS1 recorded a reversal in the bulk velocity orientations of two distinct electron populations traveling along the magnetic field, suggesting that MMS1 traversed the mid-plane (the M-N plane passing through the EDR) over the .15 s separating Figure 5b's panel (i) from panel (vi). A $B_N$ ~ 0 component of the magnetic field, present for the majority of Figure 5a, should also reside within the mid-plane. A video included via hyperlink in the supplementary information further depicts all four spacecrafts' simultaneous electron PSD measurements for Event 2, evolving in time, with other corresponding data provided. All four MMS spacecraft captured distinctly opposing populations of magnetospheric and magnetosheath electrons reverse their respective velocities relative to the local B-field direction multiple times, as the spacecraft repeatedly traversed the mid-plane along the magnetospheric edge of a nearly stationary EDR.

The nature of the large amplitude electrostatic waves is now more closely examined. Of special note in Figure 5a, we have focused on a 30 ms slice of E-field wave activity in panel (i) to analyze a pulse of electrostatic oscillations recorded during our interval of interest. The mostly-quiescent E-field of the first ~10 ms was interrupted by a soliton-like wave envelope, which also appears in the spectrogram of panel (ii) as a burst of relatively broadband, high amplitude activity (dark red). At the same instant (dashed line), $B_L$ and $B_M$ were of equal magnitude and opposite sign, a relationship roughly shared by the strong E-field waves, indicating oscillations of **E** at slight angles to the B-field. A small $E_N$ component was also present. In Figure 5c(i) and (ii), we continue Figure 5a(i) for two additional 30 ms plots, showing two additional soliton-like waves before the E-field oscillations settled back down to near-quiescent levels (1 to 10 mV/m amplitudes). In panel (i) of Figure 5b, MMS1 began on a field line connected to the Earth's north pole (hotter magnetosphere population traveling anti-parallel to **B**, with a colder magnetosheath population traveling parallel to **B**), and ended on a magnetic field line connected to the south pole, as depicted by the two populations switching sides over the following five electron PSDs. Crescents oriented parallel to the magnetic field direction, similar to those found by *Burch et al.* [2016], formed within ~15 ms of the first large wave group's passing. This sequence of observations may constitute direct evidence of electron (spatial) density enhancements and holes forming and/or propagating along the separatrix [*Khotyaintsev et al.*, 2010, *Cattell et al.*, 2005], perhaps associated with magnetic field lines breaking and/or anomalous resistivity. Electrostatic whistler waveforms reported on by *Kellogg et al.* [2010] exhibited additional perturbations superimposed on the main oscillation mode, similar to the departure from sinusoidal-like behavior shown here. The smaller amplitude superimposed fluctuations were revealed to be trapped sheets of electrons propagating along the magnetic field, propelled by the E-field perturbations of the whistlers. Although *Kellogg et al.'s*

[2010] paper characterized plasmaspheric electrostatic whistlers, MMS1 likely saw a similar type of electron transport mechanism distorting a pure sinusoidal waveform. We note that MMS's 30 ms electron sensing resolution is not high enough to directly resolve any distinct electron sheets propagating under these conditions. Additional analysis to constrain these waves' dispersion relations is already underway, in tandem with more closely constraining the path of the spacecraft during and near this event. Even though the |**j** * **E**'| values were not significantly large at the exact time of the waves we have remarked on here, a higher cadence sampling of particles (not shown) reveals cancelations in the positive and negative components of **j** * **E**' masked inside of the depicted 30 ms resolution.

**3.5) Event "3"**

After Event 2, the spacecraft made a retreat Earthwards, and then re-approached the EDR from the magnetosphere again, ~40 s later, shown in **Figure 6**. Along with continued agreement between MMS1 and MMS4 (followed by MMS3, and lastly, MMS2, though they are not shown here), other similarities exist between this event (designated Event "3") and Event 2, despite their separation in time. In Figure 6a, we have chosen four electron PSDs that depict rough qualitative and quantitative agreement with Figure 5b for electron PSD characteristics and associated $n_e$ and $\sqrt{Q_e}$ values. Figure 6b shows that MMS1's B-field, currents, baseline agyrotropy, higher-density onset, and dominance of $T_{e\parallel}$ were also all relatively close to those detailed in Figure 5a of the previous section. The E-field data recorded here shows significant sustained $+E_N$ (several 10's of ms). This $+E_N$ is the electric field believed to accelerate the electrons composing the energized crescent populations [*Shay et al.*, 2016], which we show here to exist concurrently. MMS1's electron PSDs very closely match those collected by MMS4 (Figure 6c), although some of the similarities between MMS1 and MMS4 may be attributable to a near-synchronous particle sampling, an offset of only 4 ms. Greater differences were seen between a single spacecraft's 30 ms snapshots of the EDR electrons (separated in time) than differences observed simultaneously between the two spacecraft. MMS1 and MMS4's values for **j** * **E**' (Figure 6d) were ~1/2 of those recorded during Event 2's, and neither is $T_{e\parallel}$ here as high as Event 2's, but overall, the trends suggest at least some consistency between magnetosphere EDR characteristics for similar reconnection conditions. An absence of large amplitude E-field waves might mean a reduced reconnection rate or simply a less direct EDR encounter for Event 3 compared to that of Event 2; More work is required to distinguish between these two scenarios. Very soon after the interval of time shown here, MMS returned to the sheath and continued along its outbound orbit away from the magnetopause and EDR. Although no magnetopause crossing took place during our designated window of time for Event 3, another multi-spacecraft analysis similar to our treatment of Event 1 (using temporal differences in $T_{e\perp}$ features, not shown here) yields an Earthward magnetopause speed of ~50 km/s as $B_L$ returned towards negative values, near 07:50:37 (Figure 2).

### 3.6) Summary

The data presented for November 23$^{rd}$, 2016 show that the four MMS spacecraft first crossed a reconnection *X* line with a moderate guide field (~1/2) as the magnetopause moved outward, and then made at least two crossings of the electron stagnation region, as the magnetopause continued to move repeatedly outward and inward. Several previously observed EDR signatures were seen in these three events, including:

1. Perpendicular and parallel electron crescent distributions.

2. Hot electrons traveling along **B**.
3. Ohmic dissipation.
4. Broadband, large amplitude electrostatic wave generation, from LH to above the electron cyclotron frequency, oriented roughly parallel to **B**.
5. Sustained $+E_N$ concurrent with perpendicular crescent PSDs.

The three events spanned ~one minute, suggesting that reconnection can occur continuously over minute-long time scales. In Event 2, the four spacecraft all traversed the mid-plane within a 2 s window, as evidenced by changes in the direction of the parallel crescent distributions seen in a supplemental movie (available by email request). Simultaneous measurements were made on open and closed field lines, above and below the mid-plane, and are unprecedented for MMS and thus for any measurement of reconnection in space. The large amplitude electrostatic waves seen during Event 2 exhibit possible signs of electron transport, justifying further in-depth investigation.

## 4. Overview of Thirty-Two EDR Events from Phase 1

All MMS burst data from Phase 1 were examined for occurrences of hot electrons, low |**B**|, DC or fluctuating E-field (waves), ion jet reversals, and large $+j_y$ (current density in the positive Geocentric Solar Magnetospheric y-direction), most often with indications of a magnetospheric B-field before, during, or after each interval. After narrowing down the resulting field of candidates with our criteria, several tens of thousands of electron PSDs were then searched through for their likeness to agyrotropic, two-fluid, crescent-like shapes. Very rarely do clear signatures of two distinct electron plasma populations (such as crescents) appear in the

distributions.

We now present one 3D electron PSD from each of the EDR or near-EDR encounters revealed by our analysis. Though we do not claim it to be comprehensive, the list includes thirty-two total EDR events and candidate events. **Figures 7a, 7b, & 7c** show the result, with all events sorted chronologically from left to right, then broken up into successive rows. A fully-labeled plot on the right side of the final row of each section defines the three views. The choice of timing and spacecraft for each event's sample in Figure 7 was made to be loosely representative of the most crescent-like distributions recorded over each event interval. The October 16$^{th}$, 2015 selection, the most studied EDR event to date, is the exception to the criteria; Its PSD was selected to represent one possible example of the near-EDR region electrons, to aid in comparisons. Using each event's three plots together, we can infer that most 3D distributions shown here form a roughly hemispherical shell or belt, often compressed towards lesser values of $v_\parallel$. Our 2D crescents are the projection of this 3D shape onto a plane. Although only one example per EDR candidate event is provided here, most events contain many other 30 ms windows with crescent distributions for multiple spacecraft.

A preliminary dayside EDR statistical meta-study was also conducted. Widely-studied properties of EDRs are computed and analyzed for the thirty-two events. **Table 1a & 1b** detail the results. Events are now sorted by row, chronologically, with different computations composing the table's columns. A brief description of all computations is offered in the captions. We use a 4-second computation window, beginning two seconds prior to the event's "central time", and lasting two seconds afterwards. An event's central time is generally chosen to lie at the electron PSD time shown in Figure 7, rounded to the nearest second. Some exceptions are made in the case of familiarity with an event's more exact center. Each event computation is performed first at the individual spacecraft level, and then the four results are averaged together to obtain the final numbers. As an example, "Avg. $\sqrt{Q_e}$" refers to a calculation in which each individual spacecraft's agyrotropy observations are first averaged over the 4 second window, and then averaged again across the four spacecraft. The sole exception is an exclusion of MMS3 data during encounter B26 due to a timing accuracy error, and thus only the other three spacecraft results were averaged together. Note that we have not included values of $\sqrt{Q_e}$ in the table's computations for data points when $n_e < 5$ electrons per cubic cm. Off-diagonal values in the pressure tensor are easier to increase (relative to the diagonal's values) for a plasma of lower density and dominant $T_{e\,\|}$, a condition commonly found in the outer magnetosphere known to skew our computations. Greater statistical uncertainty is also introduced at lower $n_e$. Our restriction only affects a small minority of the events' results in the average and maximum $\sqrt{Q_e}$ columns.

The plots of **Figure 8** are examples of simple first-order/linear correlation factors computed from values listed in Table 1. All five plots use the average of $\sqrt{Q_e}$ on the horizontal axis. A generally positive correlation of the average $\sqrt{Q_e}$ with the five other measurables is expected, but additional complexities due to turbulence [*Ergun et al.*, 2017], a wide assortment of reconnection conditions, and various spacecraft trajectories relative to the EDR all logically prevent a perfect 1-to-1 match between any two of our simple indicators. We feature *Burch et al.*'s [2016] October 16th, 2015 event (A03), represented by the green datapoint in each correlation plot. The November 23$^{rd}$, 2016 events are also featured, for easy comparison: B20 (Event "1"), B21 (Event "2"), and B22 (Event "3"), in yellow, magenta, and orange colors, respectively. We can see that these four events rank within the top seven highest average agyrotropy measurements. B21 contains especially large **j** * **E**' statistics. The comparisons here only serve as a crude example, but additional approaches are planned.

Although linking features and qualities of these separate events beyond elementary associations is difficult, we may now claim that MMS has observed definitive electron signatures of reconnection over a significant span of plasma parameter space. Defining a more methodical approach to EDR meta-analysis computation interval selection, requisite computables, and correlation algorithms is left for a future study. MMS data refinement is scheduled to extend perhaps several years into the future, and will offer increasingly accurate data products.

## 5. Conclusions

Using the November 23rd, 2016 events, we have demonstrated further evidence of several measures used to gauge EDR activity, including **j** * **E**', agyrotropy, crescent-shaped velocity distributions, electron heating, large **j**, and low-frequency (~10 Hz) waves across multiple quantities. During the magnetopause EDR encounter, Event 1 (in Table 1 as B20), distinctly different PSDs were seen between spacecraft, and significant gradients in several measurements occurred throughout a handful of intervals. The amount of disagreement between spacecraft is often (to first order) a function of relative spacecraft positions projected onto the N-axis of a boundary-normal coordinate system. Powerful electrostatic waves were observed on the magnetospheric side of the EDR, as rapidly changing electron behavior transpired. Waveforms similar to electrostatic whistlers known to contain sheets of trapped electrons were present. A more thorough examination is already underway.

We presented thirty-two total EDR events or strong candidates, the majority of which are listed here for the first time. Nominal mission success for the dayside phase of MMS was contingent on sixteen EDR encounters, a number we claim to have now surpassed. Our collection of encounters illustrates the variance of plasma conditions under which MMS has observed electron diffusion. In addition to discussing some general and established EDR characteristics, one goal of this study is to initiate a dialog surrounding appropriate methods of MMS EDR meta-analyses. Another aim is to continue populating the manifold of possible PSD configurations exhibited by electron distributions surrounding and inside of the EDR. Associated E-field activity and B-field topologies are of high interest. Additional analysis is required.

**Acknowledgements:**

We thank the MMS instrument team leaders for working to achieve MMS's high-resolution, multi-spacecraft data: R. Nakamura for the Active Spacecraft Potential Control (ASPOC), B. H. Mauk for the Energetic Particle Instrument (EPI) data, and R. J. Strangeway for the magnetic field data. Contributions from the entire MMS Science Working Team (SWT) aided this study. The SPEDAS (Space Physics Environment Data Analysis System) software package, used for data querying, parsing, and visualization, greatly aided in expediting our research. SPEDAS is available for download here: http://themis.ssl.berkeley.edu/socware/bleeding_edge/. The authors are also grateful to the MMS Science Data Center (SDC) of the Laboratory for Atmospheric and Space Physics (LASP) for hosting the large database of MMS observations, found here: https://lasp.colorado.edu/mms/sdc/public/links/. We thank CDAWeb data provider J. H. King, N. Papatashvilli at AdnetSystems, and NASA GSFC for the OMNI data.

The author also thanks his fiancée, Paige Nielsen.

This study was supported by NASA under grant NNX14AN55G & NNG04EB99C.

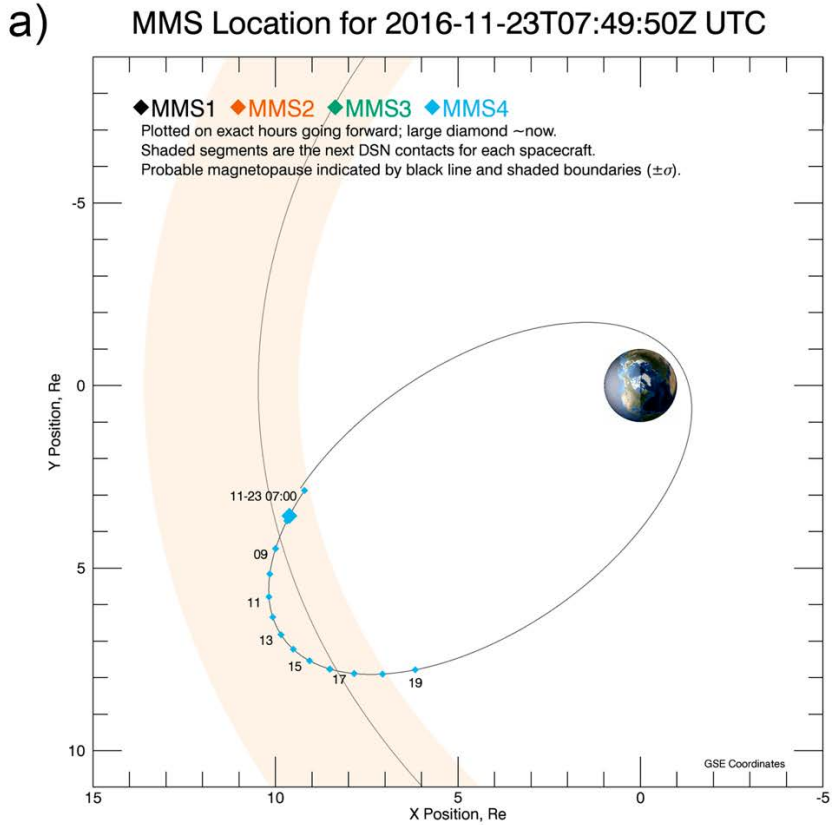

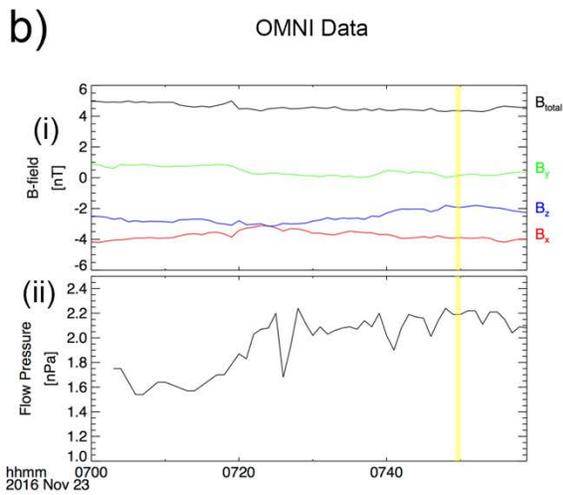
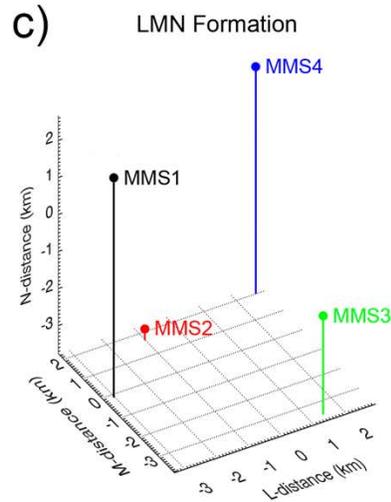

**Figure 1:** November 23$^{rd}$, 2016 event conditions. **a)** MMS's location relative to Earth and the average magnetopause boundary, shown in GSE, in units of Earth radii. **b)** Beginning at 07:00 UTC, a 1 hour plot of the solar wind conditions. Panel (i) plots the IMF magnitude, in black, and the x, y, and z (GSM) components in red, green, and blue, respectively. Panel (ii) shows dynamic/ram pressure. The highlighted sub-interval designates 07:49 - :50, the approximate timing leading up to the November 23$^{rd}$, 2016 EDR events. **c)** Relative spacecraft positions in LMN coordinates (see Sect. 3.1), in km. The origin is placed at the constellation's centroid. MMS1 is black, MMS2 is red, MMS3 is green, and MMS4 is blue.

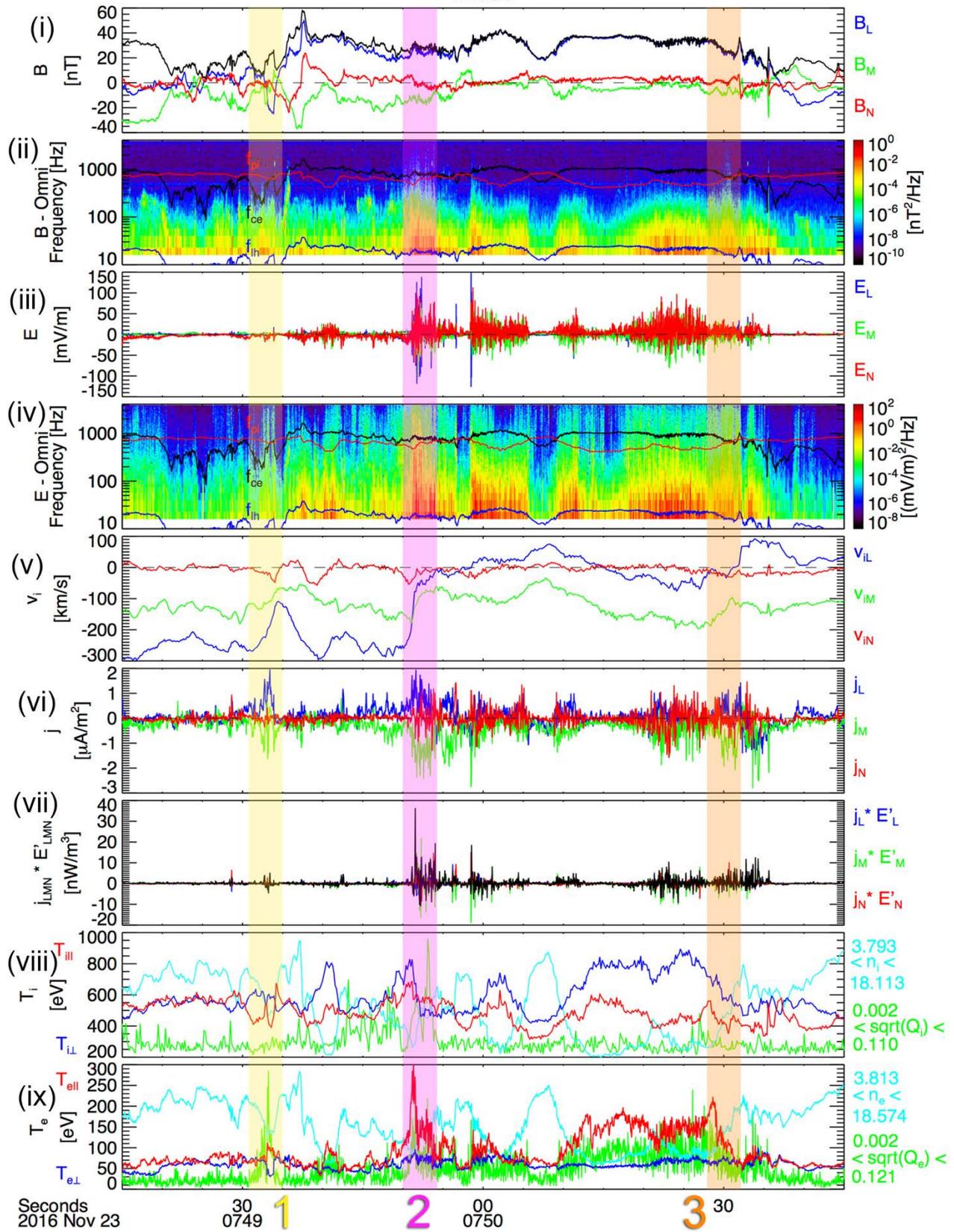

**Figure 2:** MMS3's overview of the EDR observations on November 23[rd], 2016. Vector components are given in LMN coordinates (L=blue, M=green, N=red), and black traces indicate a vector's total magnitude. Panel (i) shows the B-field, and (ii), the B-field waves spectrogram. Similarly, (iii) shows the E-field, and (iv), the E-field waves spectrogram. The spectrograms also show the computed frequencies of $f_{lh}$ (lower hybrid - blue), $f_{ce}$ (electron cyclotron - black), and $f_{pi}$ (ion plasma - red). Panel (v) plots the ion velocity, (vi) is **j**, (vii) is **j** * **E**', and (viii) shows the ion temperature components relative to the local magnetic field, plotted with the ion density ($n_i$) and ion agyrotropy ($\sqrt{Q_i}$) scaled to lie within the panel. The maximum and minimum values of $n_i$ and $\sqrt{Q_i}$ within the plotted timespan are listed to the right of the panel, to aid in interpretation. Panel (ix) keeps the same convention used for (viii), applied to the electrons. EDR event timings are indicated by a number and corresponding color (1=yellow, 2=magenta, 3=orange).

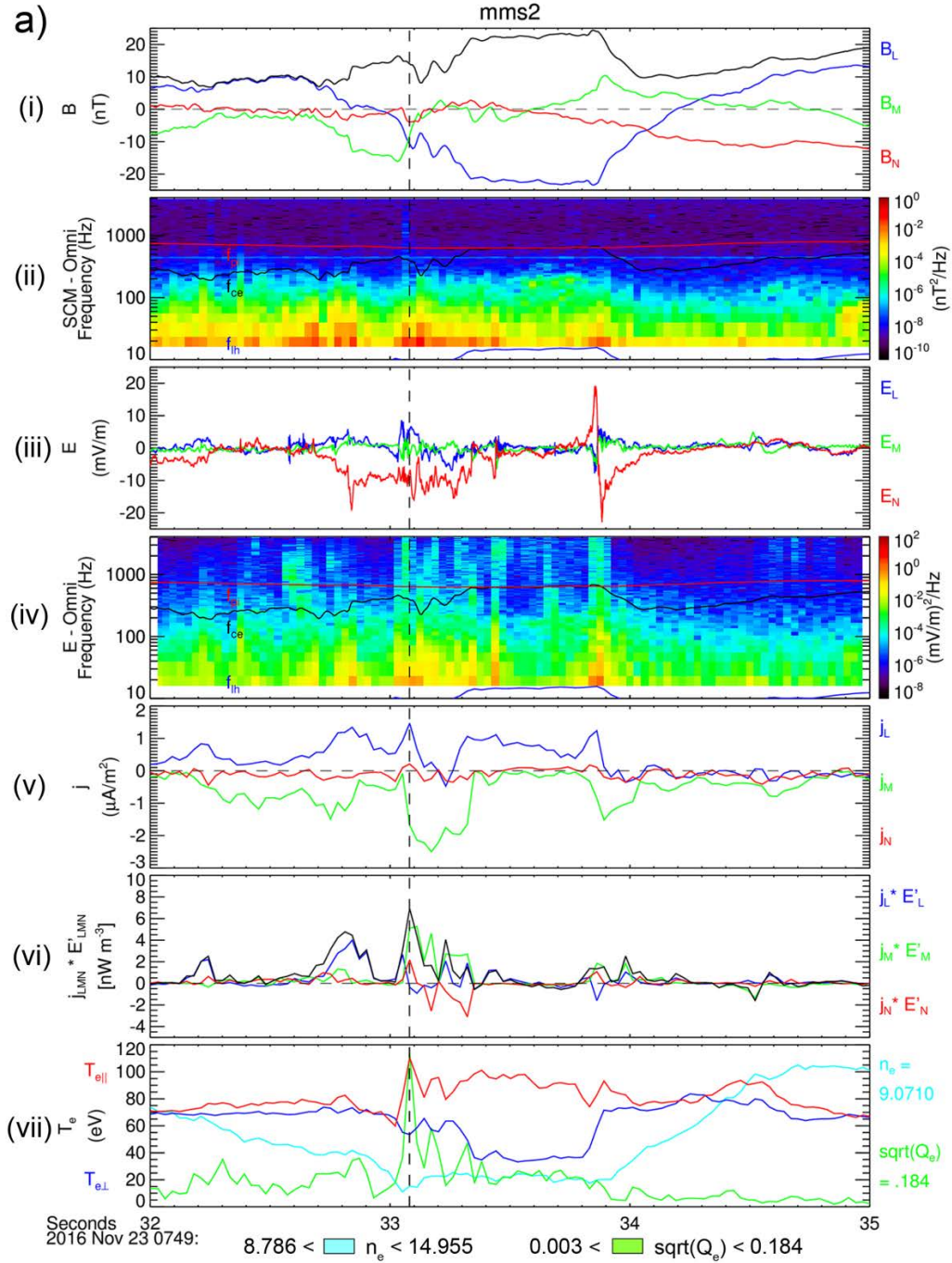

**Figure 3a):** A closer look at Event 1, the first of three EDR encounters, showing MMS2's observations for a 3 s subset of Figure 2. Panels (i) through (iv) correspond to the B-field, B-field spectrogram, E-field, and E-field spectrogram, respectively. As in Figure 2, the spectrograms show the same characteristic plasma frequencies as line traces. We graph **j** in panel (v), **j** * **E**' (broken into LMN components) in (vi), and the electron temperature, density and agyrotropy in panel (vii). The values of density and agyrotropy to the right of panel (vii) occur at the dotted line, with the minimum and maximum of each provided below the plot. All vectors are again shown in boundary-normal coordinates, with magnitudes indicated by a solid black trace.

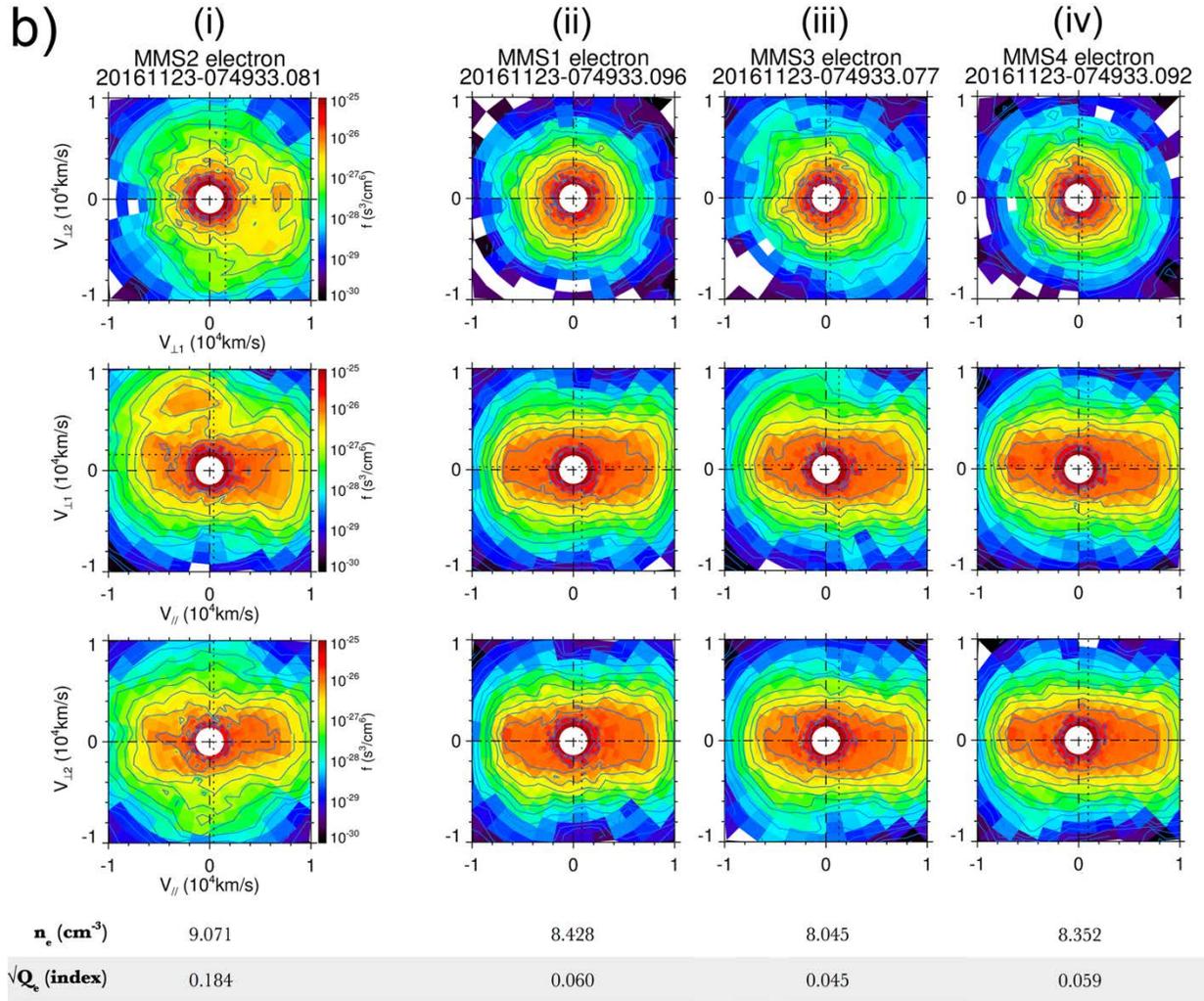

**Figure 3b):** We compare four near-simultaneous electron PSDs, one from each spacecraft, with each PSD's electron density and agyrotropy listed below. The corresponding time of the MMS2 distribution in panel (i) is indicated by the vertical dashed line in **a)**.

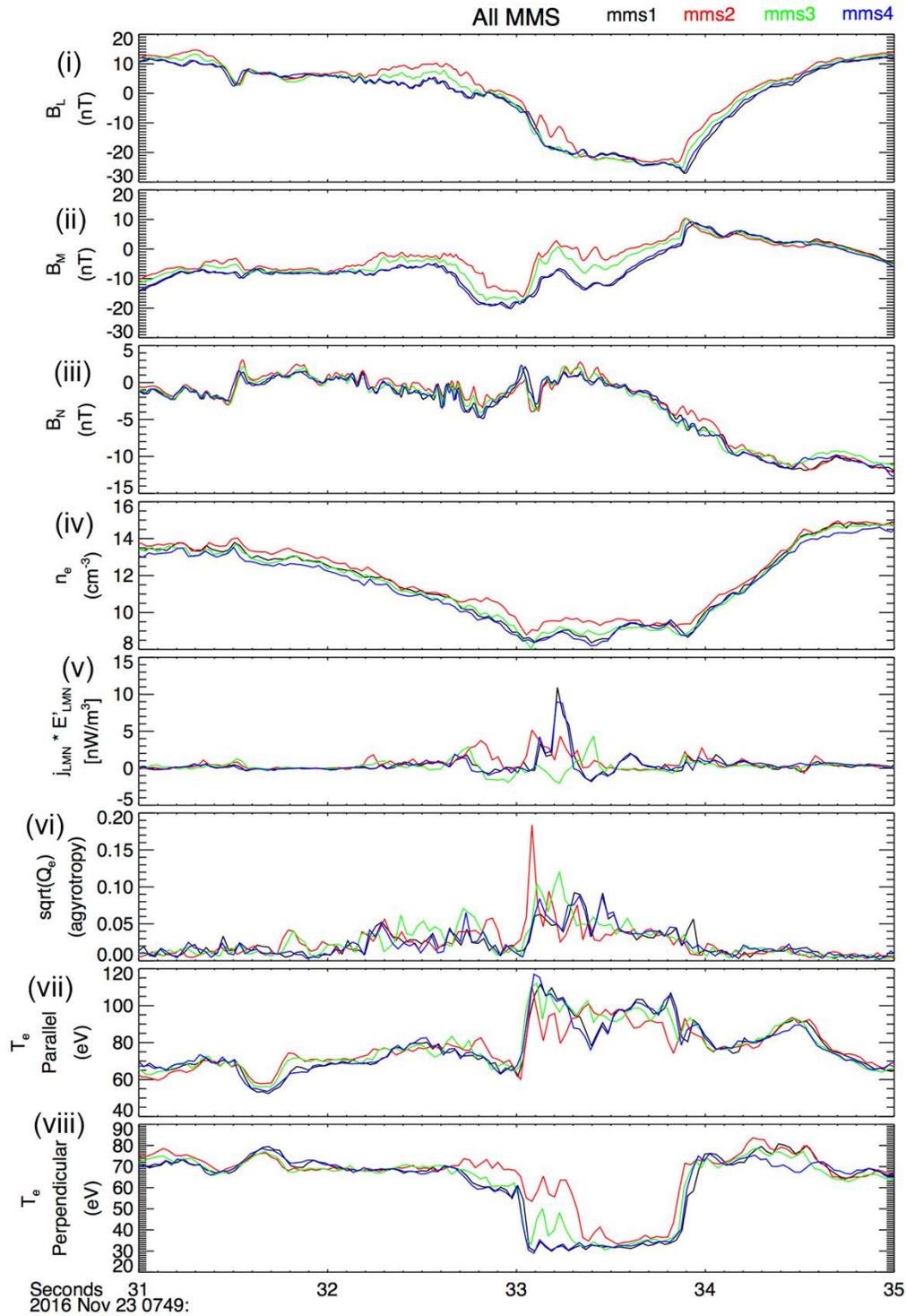

**Figure 4:** Multi-spacecraft analysis of Event 1. MMS1 is black, MMS2 is red, MMS3 is green, and MMS4 is blue. Panel (i) shows the L-component of B, (ii) is the M-component, and (iii), the N-component. Panel (iv) is electron number density. In (v), we show **j** * **E'**, and in (vi), $\sqrt{Q_e}$. The electron temperatures parallel to B are given in panel (vii), and (viii) gives the perpendicular $T_e$ component.

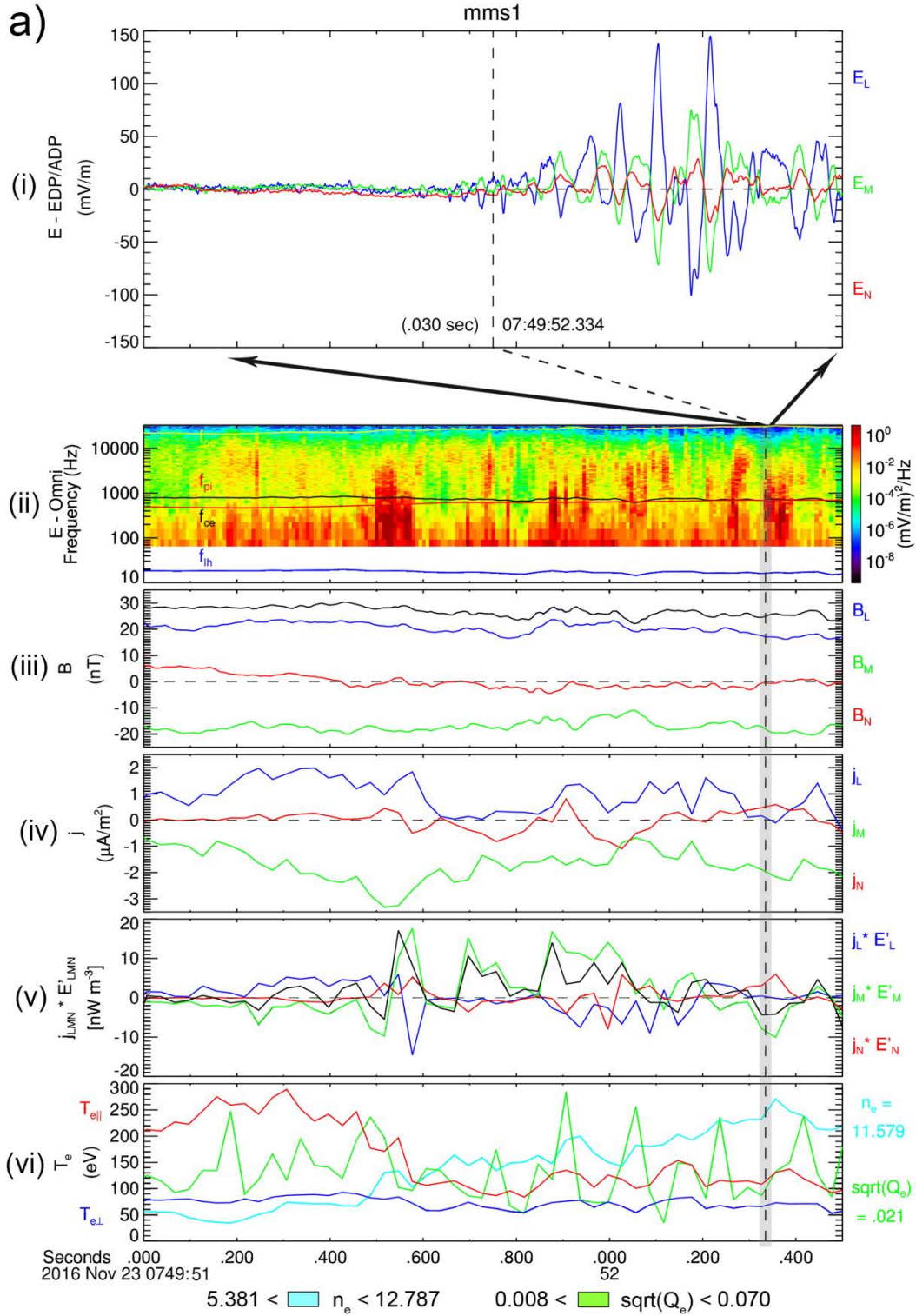

**Figure 5a):** MMS1 data from Event 2. Panel (i) shows 30 ms of high-frequency E-field wave data centered around the dashed line, which is drawn in at the same time for the other panels. The dashed line's exact time is also printed in panel (i). Panel (ii) plots the E-field spectrogram, (iii) is the components and magnitude of the B-field, (iv) is **j**, (v) is **j** * **E**', and (vi) is the same electron temperature, density, and agyrotropy scheme used in panel (vii) of Figure 3a.

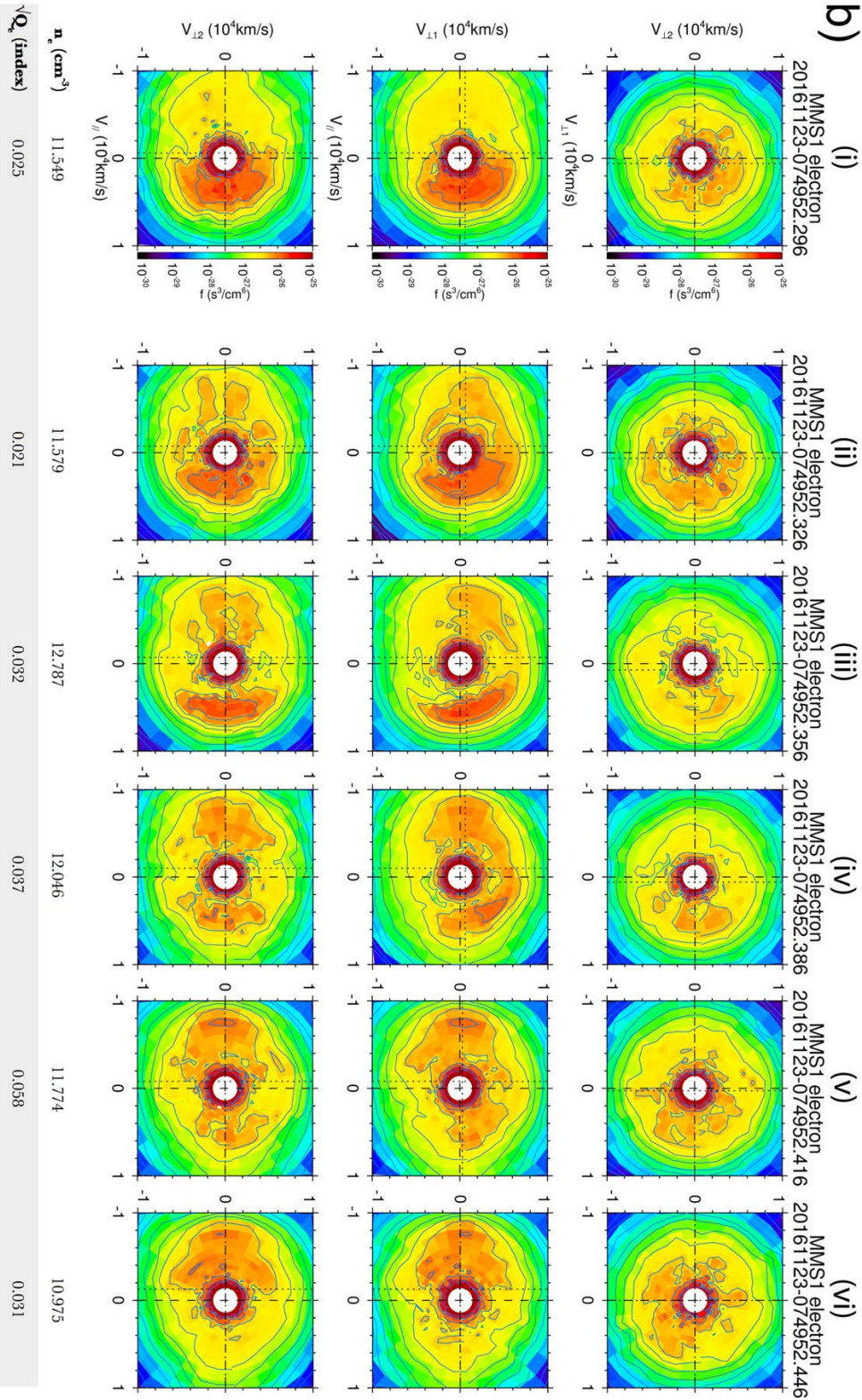

**Figure 5b):** Several electron distributions from MMS1, with electron number densities and agyrotropies below. The dashed vertical line of **a)** occurs 23 ms into the 30 ms window of time over which the distribution shown in panel (ii) of **c)** is accumulated.

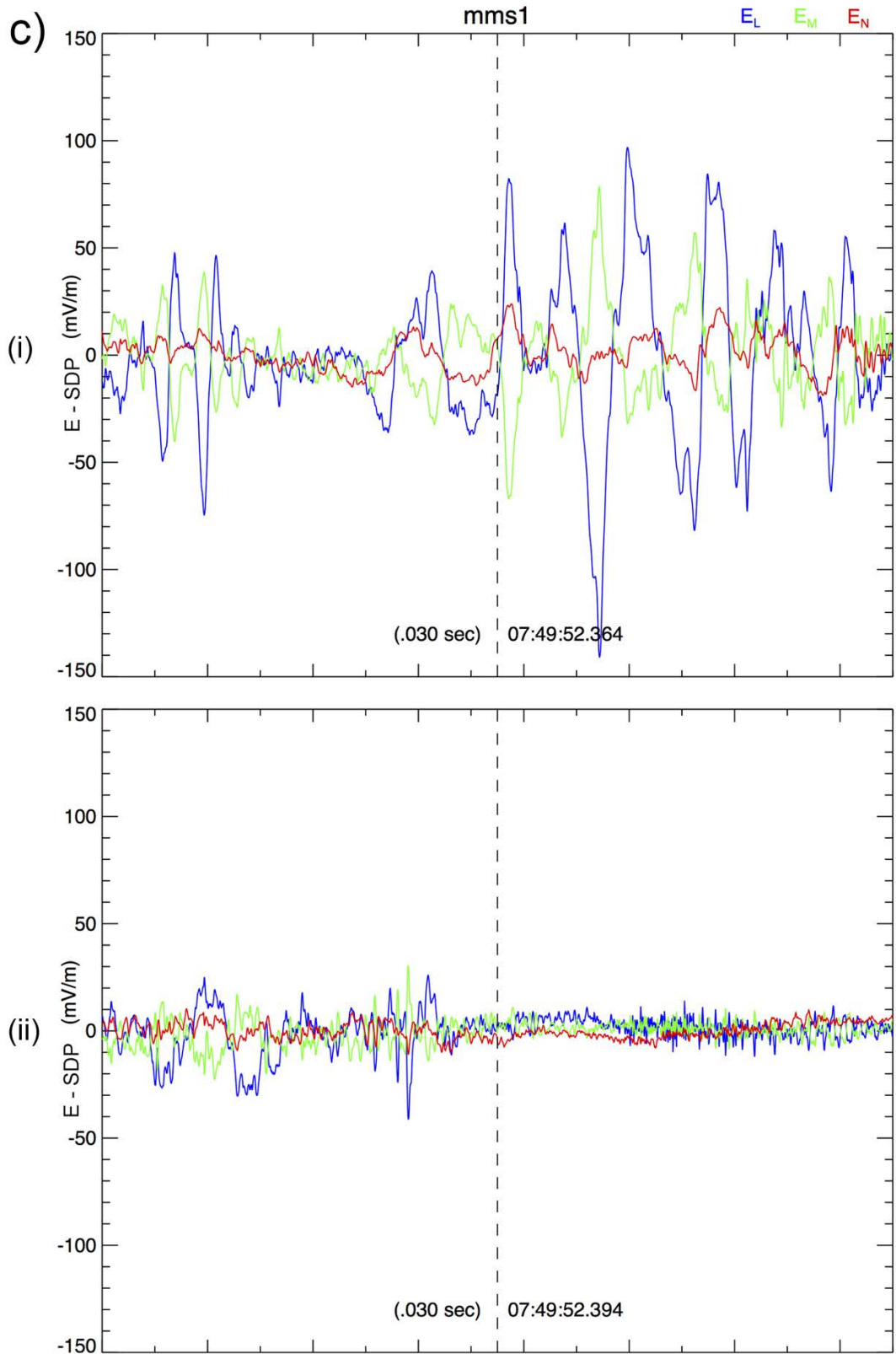

**Figure 5c):** Panel (i) and (ii) are the first and second halves of the next 60 ms of MMS1 E-field data, following the 30 ms span of E-field data shown in panel (i) of **a)**.

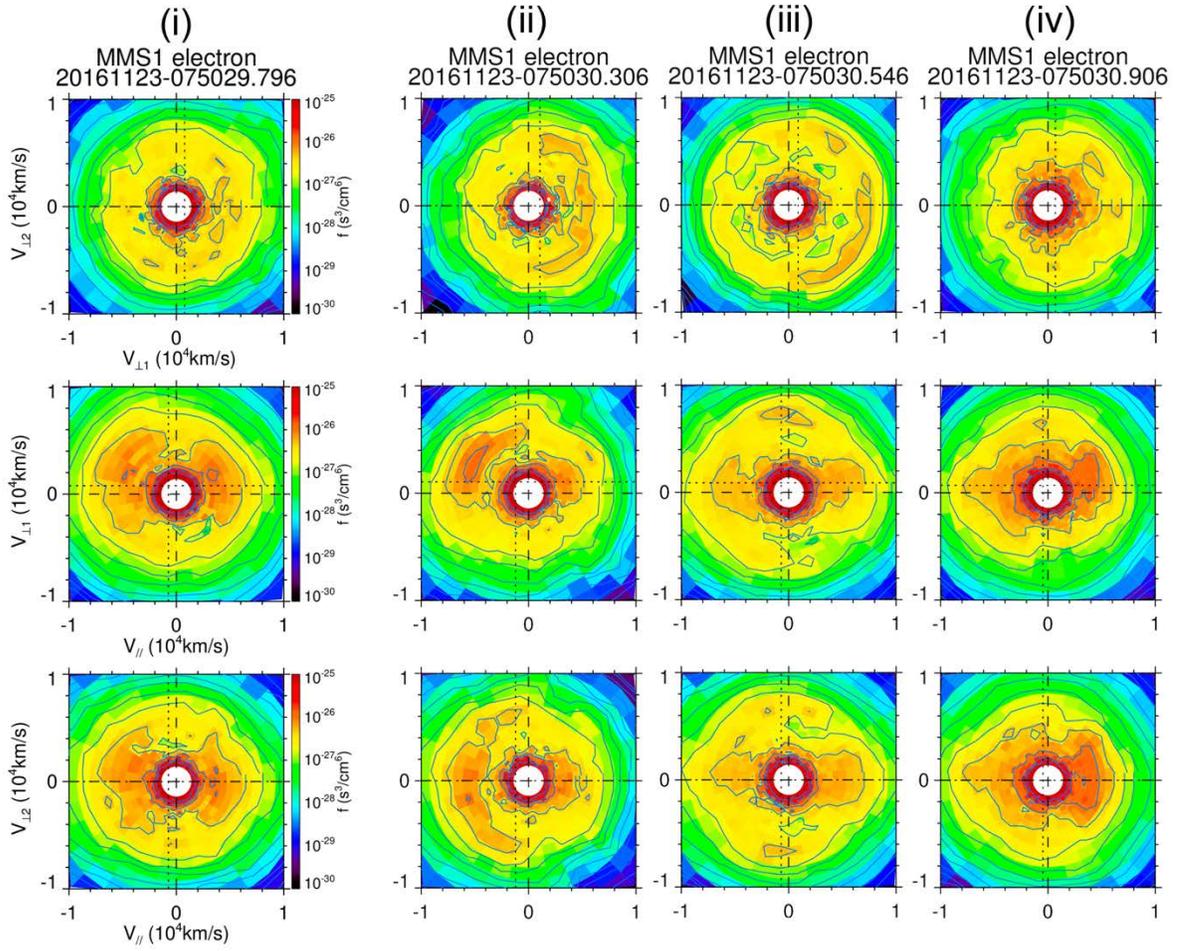

**Figure 6a):** Four electron PSDs from MMS1 are shown, again, with the respective values of $n_e$ and $\sqrt{Q_e}$ from each's 30 ms interval.

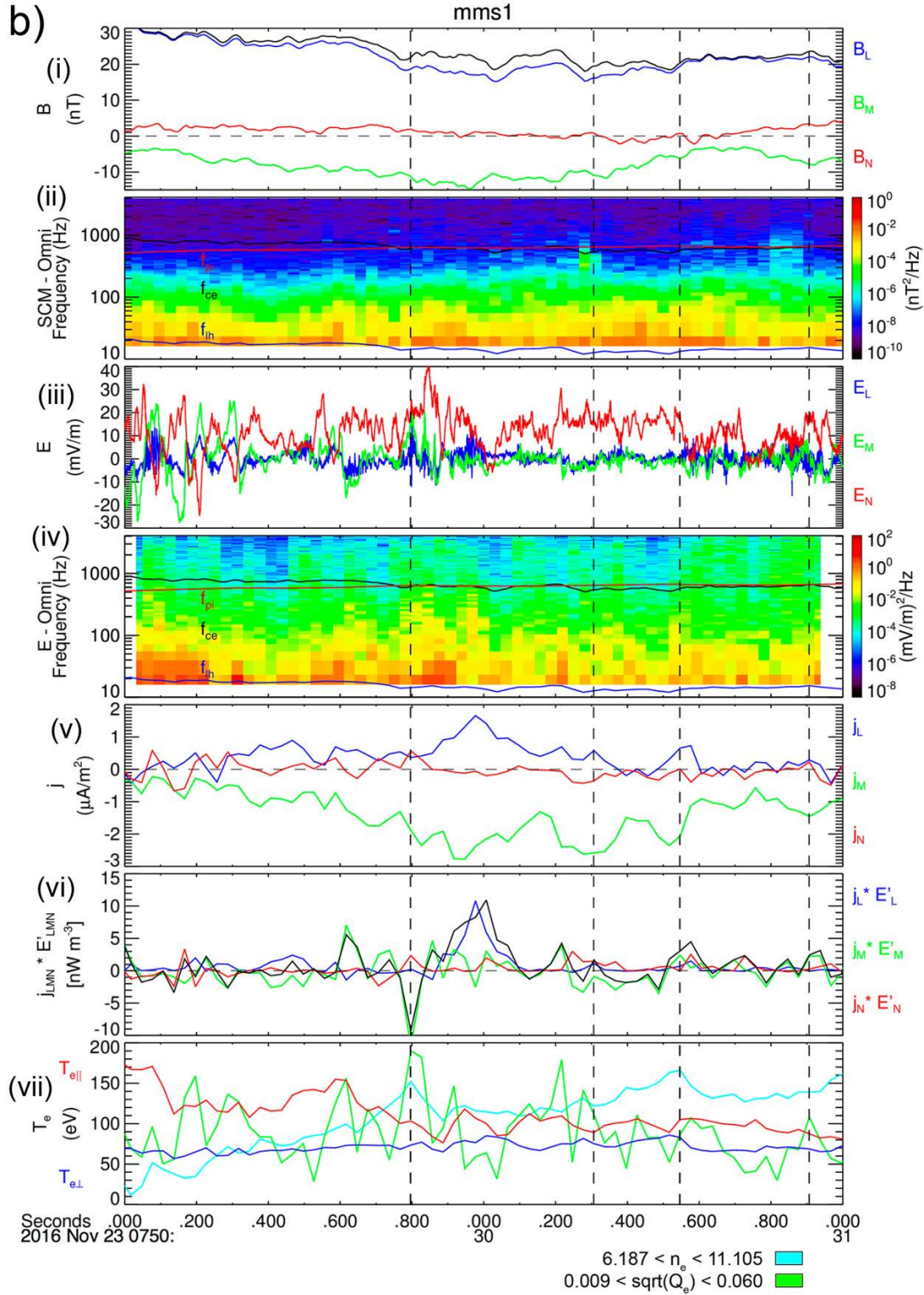

**Figure 6b):** A 2 s duration plot of MMS1 data is displayed. Panels are: (i) - B-field, (ii) - B-field spectrogram, (iii) - E-field, (iv) - E-field spectrogram, (v) - **j**, (vi) - **j * E'**, and (vii) - $T_e$, $n_e$, and $\sqrt{Q_e}$. The minimum and maximum values of $n_e$ and agyrotropy over the plot's 2 seconds are listed at the bottom. Dashed vertical lines indicate electron distribution timings for the plots of **a)**.

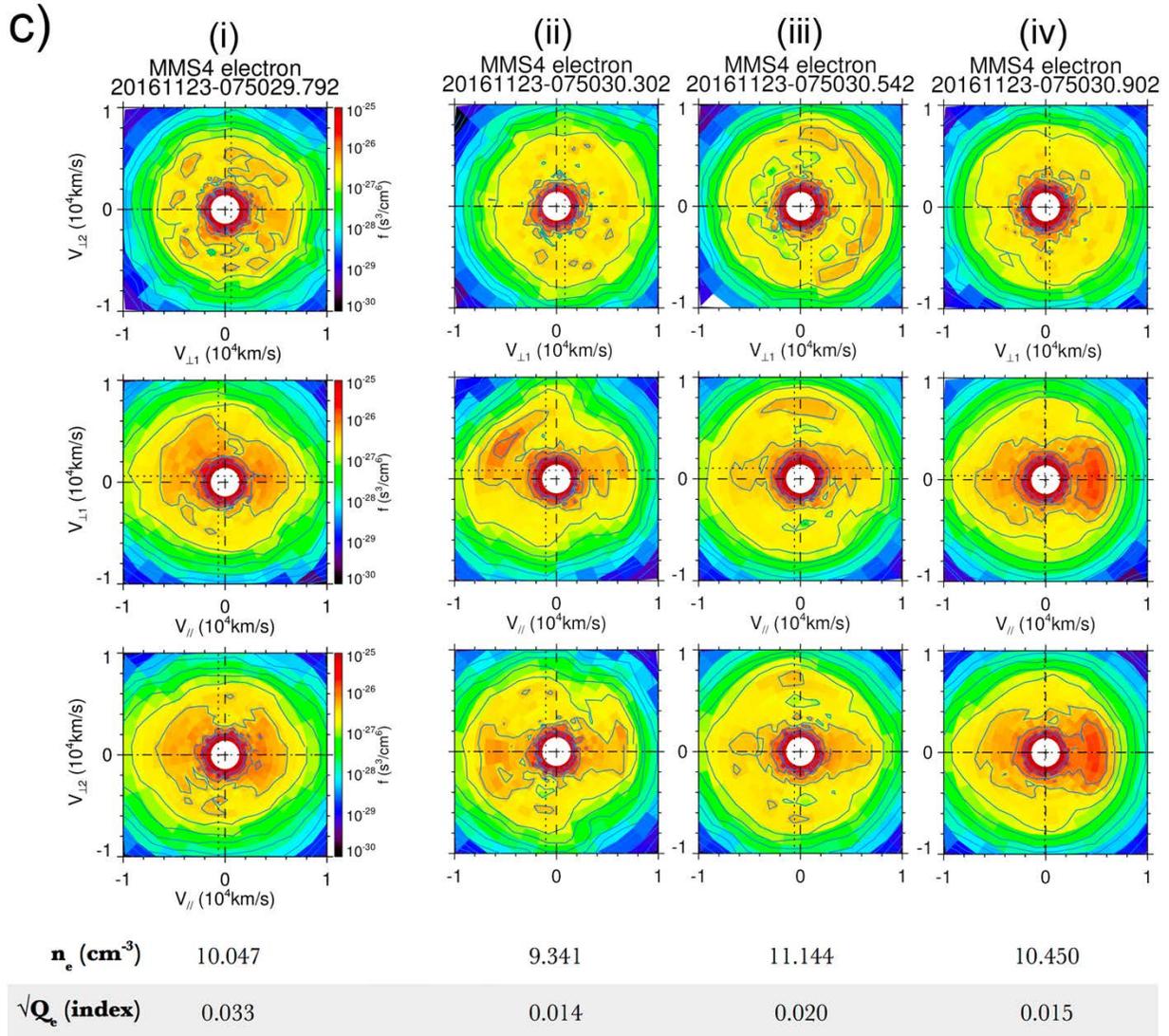

**Figure 6c):** MMS4 data is presented in an identical format to **a).**

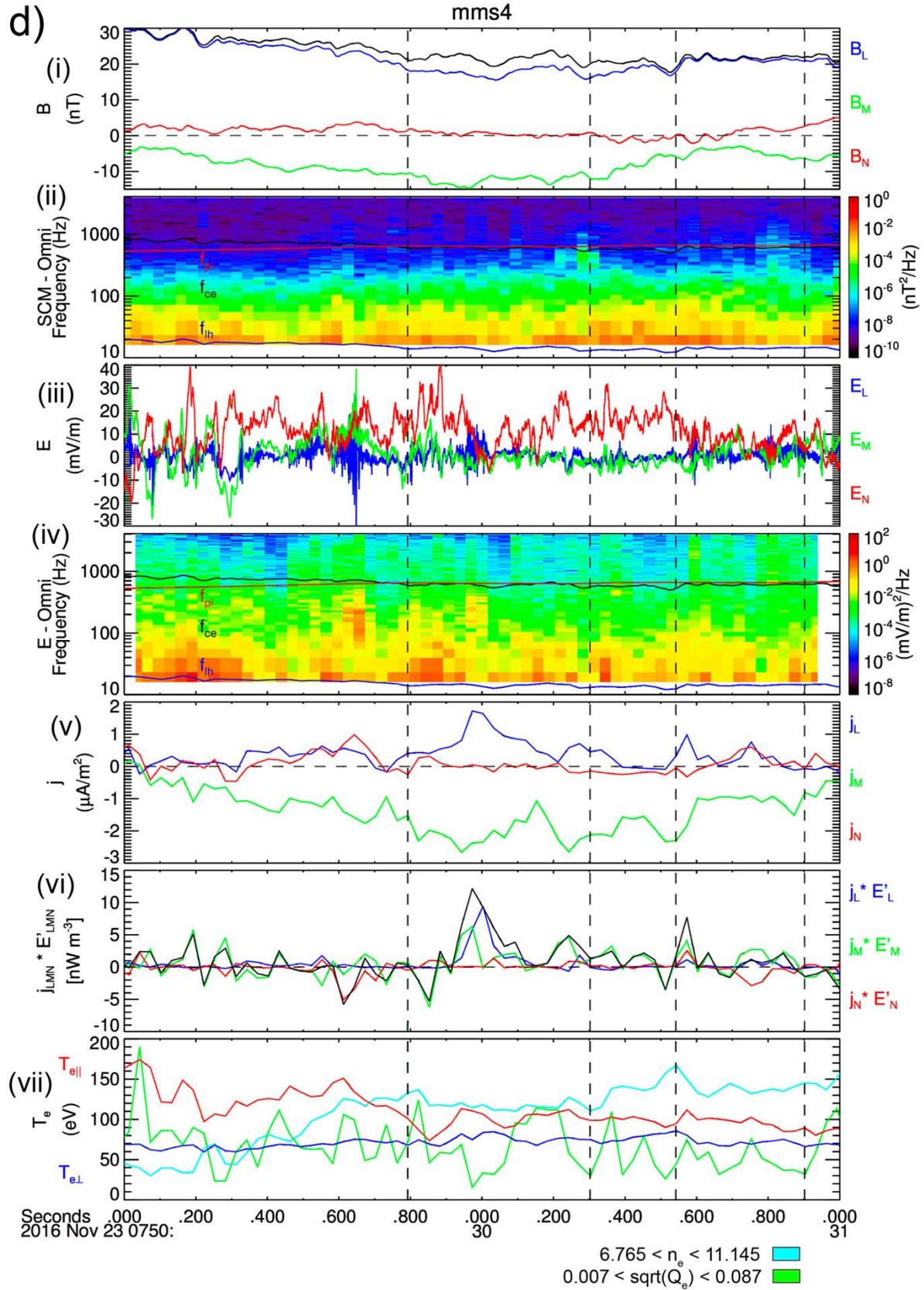

**Figure 6d):** MMS4 data is presented in an identical format to **b)**.

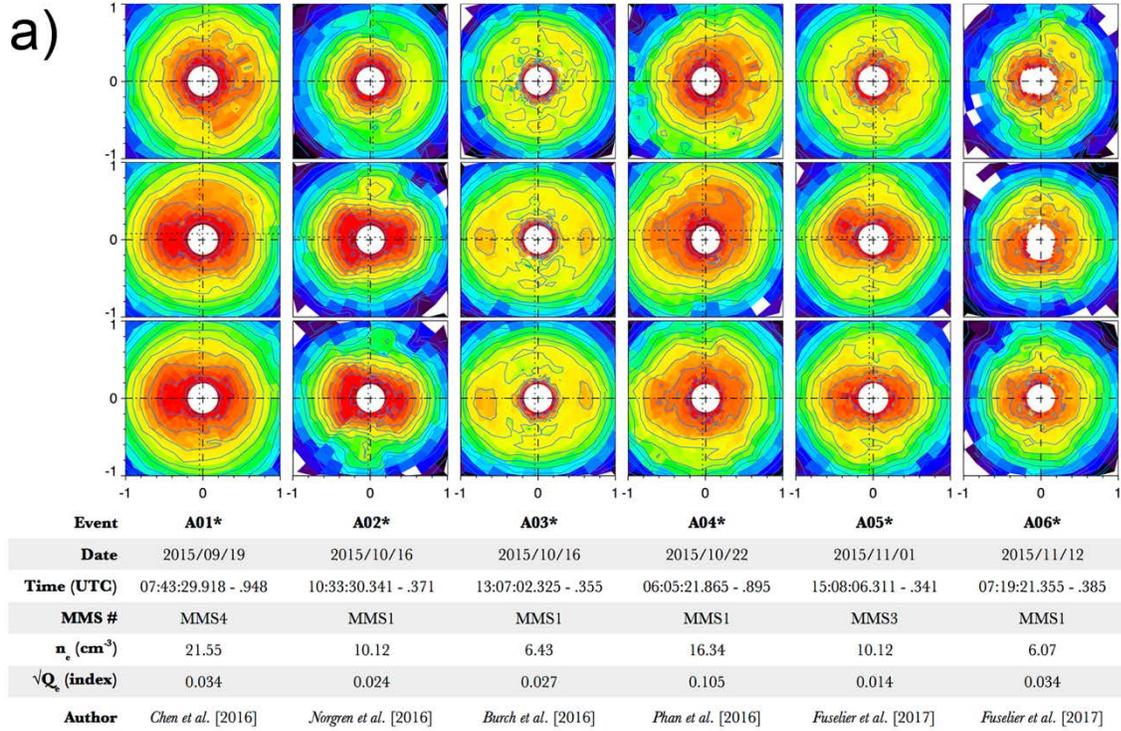
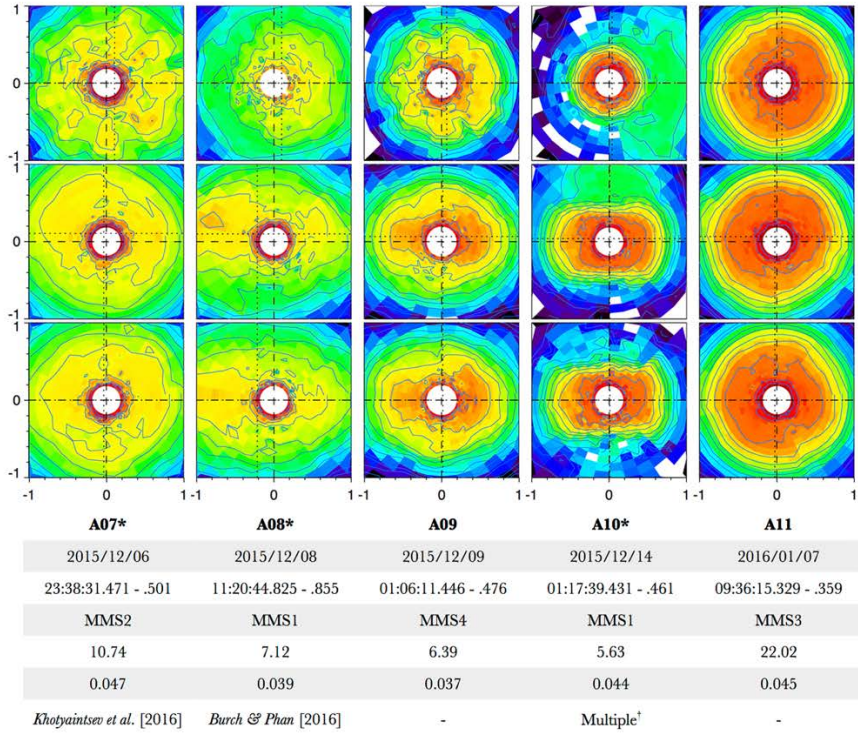
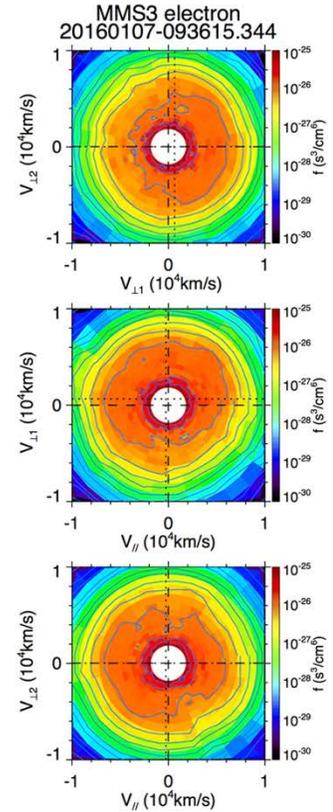

**Figure 7a):** Three views of one spacecraft's 3D electron velocity-space distribution for each of thirty-two events, sorted chronologically, first by column, then by row. Below each column, we assign an event name and list the interval of time over which each distribution function was accumulated. Names beginning with "A" designate Phase 1a events, and "B" designates Phase 1b events. We include the spacecraft number, along with the observed density and agyrotropy calculation for the same 30 ms interval. Also shown are the definitions of the axes used in each event's three cross-sectional views, at the bottom right of the figure. The average direction of the local magnetic field during each sample defines the $+\mathbf{v}_\parallel$ direction. The positive $\mathbf{v}_{\perp 1}$ direction is defined as $(\mathbf{v}_\parallel \times \mathbf{v}_e) \times \mathbf{v}_\parallel$, where $\mathbf{v}_e$ is the bulk electron velocity's unit vector, here. The $\mathbf{v}_{\perp 2}$ direction completes a right-handed coordinate system, such that $\mathbf{v}_{\perp 1} \times \mathbf{v}_{\perp 2} = \mathbf{v}_\parallel$. See Section 4 for additional remarks.

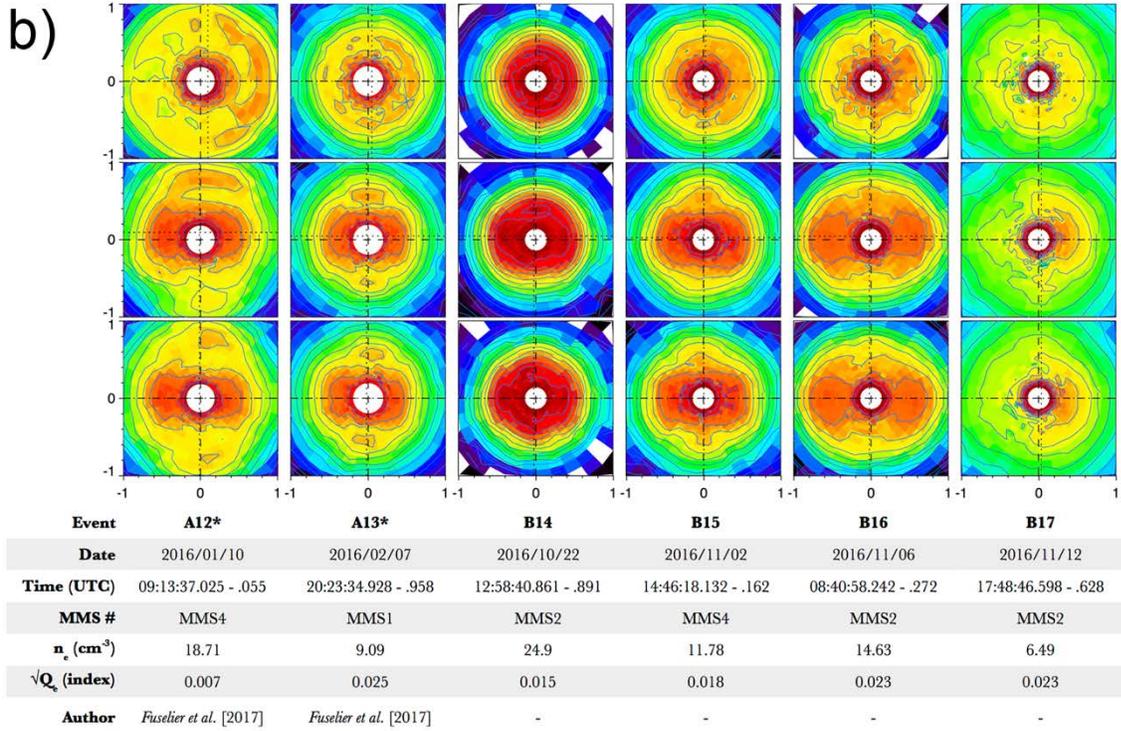
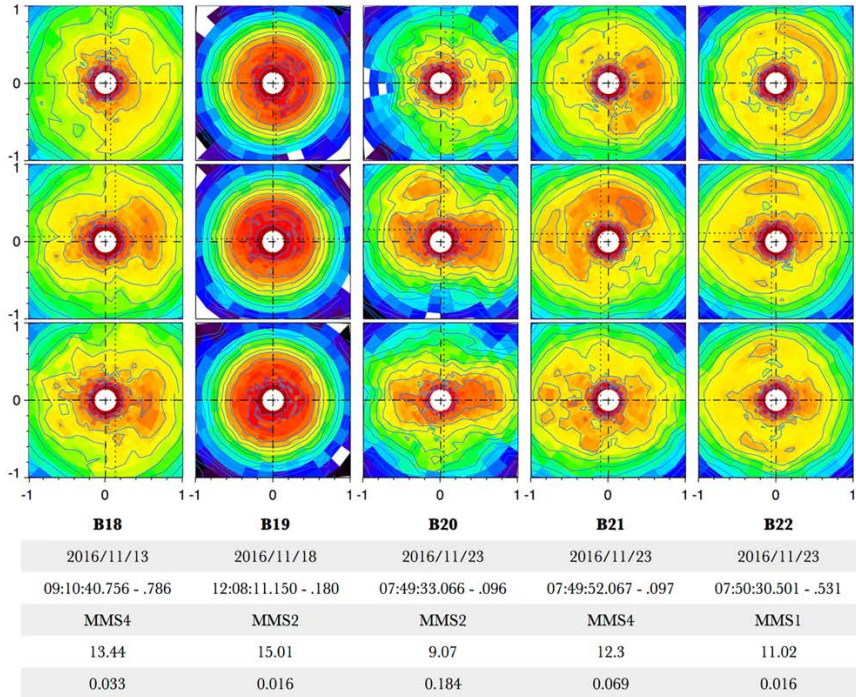
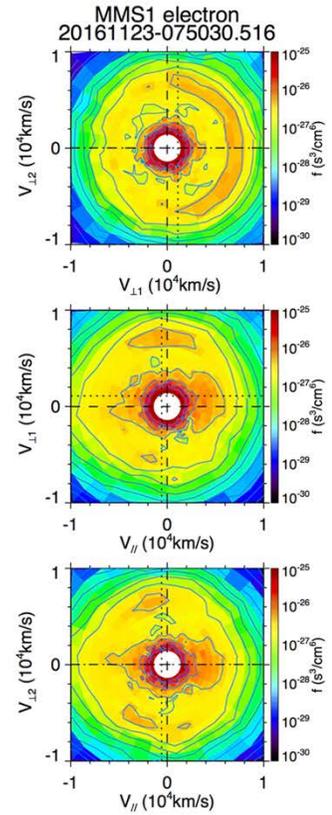

**Figure 7b):** A continuation of Figure 7a. See Section 4 for additional remarks.

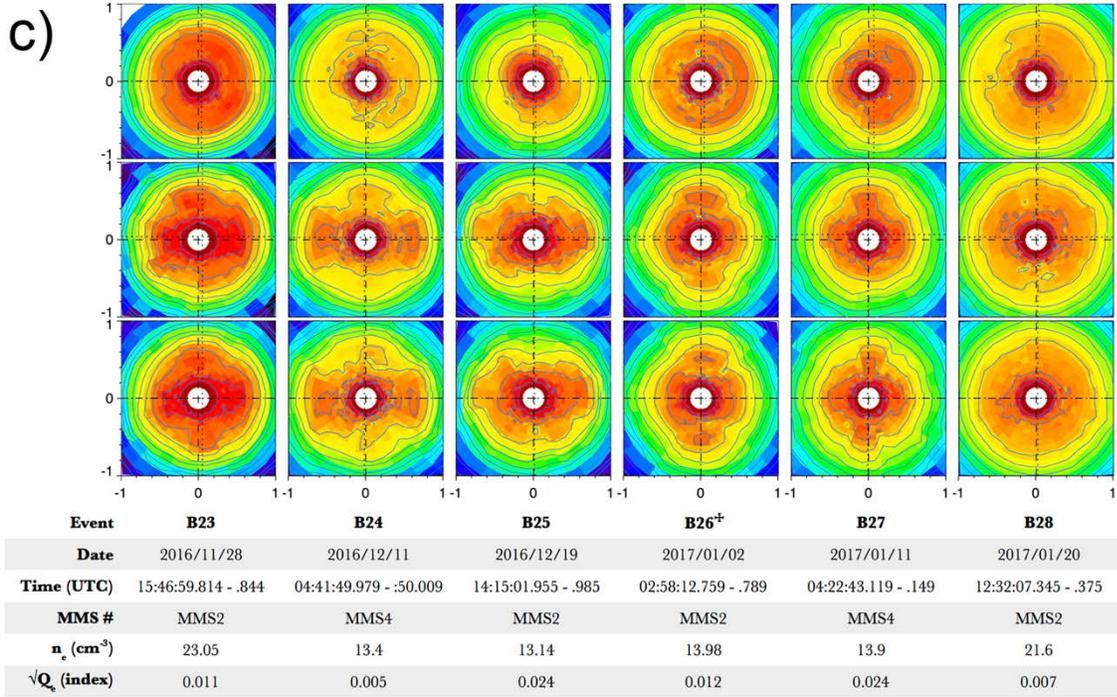

| Event | B23 | B24 | B25 | B26† | B27 | B28 |
|---|---|---|---|---|---|---|
| Date | 2016/11/28 | 2016/12/11 | 2016/12/19 | 2017/01/02 | 2017/01/11 | 2017/01/20 |
| Time (UTC) | 15:46:59.814 - .844 | 04:41:49.979 - :50.009 | 14:15:01.955 - .985 | 02:58:12.759 - .789 | 04:22:43.119 - .149 | 12:32:07.345 - .375 |
| MMS # | MMS2 | MMS4 | MMS2 | MMS2 | MMS4 | MMS2 |
| $n_e$ (cm$^{-3}$) | 23.05 | 13.4 | 13.14 | 13.98 | 13.9 | 21.6 |
| $\sqrt{Q_e}$ (index) | 0.011 | 0.005 | 0.024 | 0.012 | 0.024 | 0.007 |
| Author | - | - | - | - | - | - |

*previously reported
†*Ergun et al.* [2017], *Chen et al.* [2017], *Graham et al.* [2017b]
‡MMS3 data outage

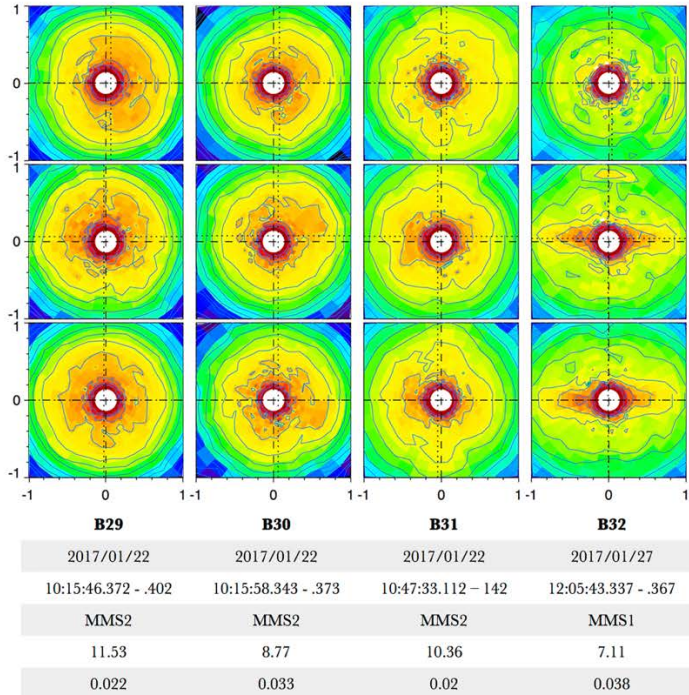
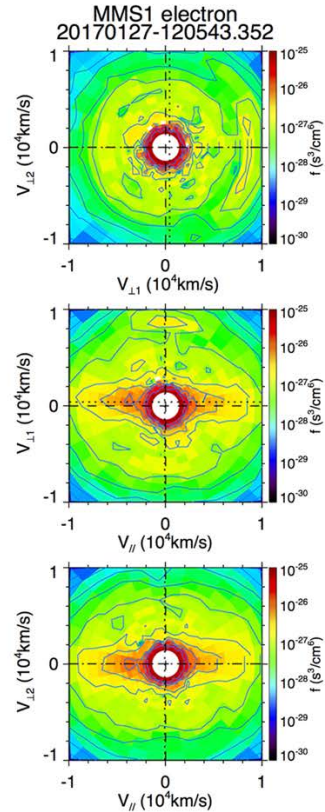

| | B29 | B30 | B31 | B32 |
|---|---|---|---|---|
| | 2017/01/22 | 2017/01/22 | 2017/01/22 | 2017/01/27 |
| | 10:15:46.372 - .402 | 10:15:58.343 - .373 | 10:47:33.112 - 142 | 12:05:43.337 - .367 |
| | MMS2 | MMS2 | MMS2 | MMS1 |
| | 11.53 | 8.77 | 10.36 | 7.11 |
| | 0.022 | 0.033 | 0.02 | 0.038 |
| | - | - | - | - |

**Figure 7c):** A continuation of Figures 7a & 7b. See Section 4 for additional remarks.

a)

| Event | Date & Time (UTC) | $X_{GSM}$ (rE) | $Y_{GSM}$ (rE) | $Z_{GSM}$ (rE) | Separation (km) | Max $\rho_e$ (km) | Max $\rho_i$ (km) | Avg. $n_e$ (cm$^{-3}$) | Avg.\|j\| (µA*m$^{-2}$) | Max \|j\| (µA*m$^{-2}$) | Max $j_y$ (µA*m$^{-2}$) | Min \|B\| (nT) |
|---|---|---|---|---|---|---|---|---|---|---|---|---|
| A01* | 2015/09/19 07:43:30 | 6.346 | 5.399 | -2.982 | 71.57 | 6.83 | 1374.9 | 22.93 | 0.79 | 2.73 | 1.84 | 10.14 |
| A02* | 2015/10/16 10:33:30 | 9.231 | 6.092 | -4.403 | 13.87 | 8.72 | 1703.8 | 13.55 | 0.56 | 1.20 | 0.90 | 2.42 |
| A03* | 2015/10/16 13:07:02 | 8.310 | 7.078 | -4.800 | 13.78 | 11.59 | 2401.1 | 6.92 | 0.63 | 1.85 | 1.52 | 2.24 |
| A04* | 2015/10/22 06:05:22 | 9.637 | 3.481 | -1.961 | 16.93 | 9.35 | 509.9 | 15.17 | 0.75 | 2.74 | 2.46 | 3.95 |
| A05* | 2015/11/01 15:08:06 | 7.814 | 6.202 | -3.470 | 14.58 | 4.25 | 112.0 | 9.12 | 0.71 | 1.98 | 1.48 | 19.49 |
| A06* | 2015/11/12 07:19:21 | 11.507 | 2.302 | -1.776 | 15.82 | 8.67 | 570.8 | 5.84 | 0.24 | 1.06 | 0.75 | 3.75 |
| A07* | 2015/12/06 23:38:31 | 8.516 | -3.916 | -0.810 | 19.23 | 5.74 | 108.3 | 9.16 | 0.69 | 2.82 | 1.61 | 19.76 |
| A08* | 2015/12/08 11:20:44 | 10.233 | 1.288 | -1.364 | 15.30 | 4.83 | 174.6 | 5.12 | 0.44 | 2.43 | 2.15 | 14.76 |
| A09 | 2015/12/09 01:06:11 | 9.922 | -3.671 | -0.928 | 17.34 | 6.10 | 619.2 | 8.03 | 0.37 | 1.11 | 0.84 | 9.85 |
| A10* | 2015/12/14 01:17:40 | 10.131 | -4.163 | -1.191 | 16.97 | 9.46 | 846.1 | 5.09 | 0.49 | 1.73 | 1.68 | 4.49 |
| A11 | 2016/01/07 09:36:15 | 8.888 | -1.968 | -0.733 | 41.75 | 7.24 | 348.7 | 21.41 | 0.64 | 2.18 | 1.93 | 8.03 |
| A12* | 2016/01/10 09:13:37 | 8.808 | -2.395 | -0.775 | 40.84 | 6.67 | 386.4 | 14.74 | 0.83 | 3.75 | 3.13 | 9.94 |
| A13* | 2016/02/07 20:23:35 | 3.874 | -9.325 | -5.720 | 15.99 | 10.72 | 2327.4 | 8.49 | 0.18 | 0.52 | 0.19 | 3.06 |
| B14 | 2016/10/22 12:58:41 | 6.406 | 7.700 | -4.706 | 8.87 | 6.21 | 331.3 | 25.49 | 0.85 | 2.81 | 2.17 | 3.81 |
| B15 | 2016/11/02 14:46:18 | 7.241 | 8.812 | -3.543 | 8.18 | 6.16 | 159.7 | 9.83 | 0.28 | 0.90 | 0.60 | 8.67 |
| B16 | 2016/11/06 08:40:58 | 7.943 | 4.113 | -2.826 | 11.76 | 3.00 | 133.1 | 13.70 | 0.96 | 2.40 | 1.82 | 28.89 |
| B17 | 2016/11/12 17:48:47 | 6.624 | 9.165 | -1.104 | 7.35 | 10.98 | 432.7 | 5.04 | 0.34 | 0.96 | 0.70 | 6.10 |
| B18 | 2016/11/13 09:10:41 | 8.958 | 4.563 | -2.625 | 11.38 | 5.61 | 341.7 | 9.40 | 0.48 | 2.92 | 1.98 | 18.57 |
| B19 | 2016/11/18 12:08:11 | 9.596 | 6.460 | -2.509 | 4.88 | 8.40 | 474.8 | 15.45 | 0.54 | 1.47 | 1.38 | 3.21 |
| B20 | 2016/11/23 07:49:33 | 9.613 | 3.232 | -1.604 | 6.43 | 8.16 | 413.9 | 11.61 | 0.73 | 2.23 | 1.56 | 5.93 |
| B21 | 2016/11/23 07:49:52 | 9.613 | 3.232 | -1.604 | 6.43 | 4.25 | 106.9 | 8.78 | 1.36 | 2.96 | 2.77 | 22.24 |
| B22 | 2016/11/23 07:50:30 | 9.620 | 3.245 | -1.608 | 6.42 | 4.66 | 110.0 | 9.05 | 1.04 | 2.59 | 2.30 | 19.90 |
| B23 | 2016/11/28 15:47:00 | 8.884 | 7.184 | -0.440 | 6.32 | 4.80 | 150.6 | 16.94 | 0.33 | 1.80 | 0.46 | 12.99 |
| B24 | 2016/12/11 04:41:50 | 9.489 | -0.056 | -0.448 | 6.89 | 7.79 | 400.0 | 13.10 | 0.35 | 0.90 | 0.74 | 8.03 |
| B25 | 2016/12/19 14:15:02 | 10.204 | 4.170 | 0.934 | 8.42 | 7.90 | 669.9 | 11.12 | 0.42 | 1.00 | 0.84 | 5.52 |
| B26† | 2017/01/02 02:58:13 | 9.647 | -3.007 | -0.649 | 9.96 | 8.50 | 761.0 | 12.71 | 0.37 | 0.89 | 0.78 | 4.30 |
| B27 | 2017/01/11 04:22:43 | 10.809 | -3.713 | -0.154 | 8.17 | 9.10 | 488.5 | 13.16 | 0.33 | 1.66 | 0.95 | 5.14 |
| B28 | 2017/01/20 12:32:07 | 9.634 | -0.461 | 1.967 | 6.47 | 9.27 | 485.1 | 16.51 | 0.72 | 1.96 | 1.25 | 6.01 |
| B29 | 2017/01/22 10:15:46 | 10.744 | -2.138 | 1.766 | 5.75 | 13.70 | 1006.4 | 10.78 | 0.45 | 1.82 | 1.63 | 2.35 |
| B30 | 2017/01/22 10:15:58 | 10.750 | -2.148 | 1.764 | 5.74 | 9.99 | 475.9 | 9.79 | 0.47 | 1.29 | 1.17 | 4.53 |
| B31 | 2017/01/22 10:47:33 | 10.519 | -1.790 | 1.837 | 5.86 | 7.48 | 391.9 | 9.59 | 0.44 | 1.17 | 0.97 | 11.43 |
| B32 | 2017/01/27 12:05:43 | 9.270 | -1.370 | 1.964 | 6.05 | 13.22 | 1054.9 | 7.27 | 0.47 | 1.49 | 1.23 | 2.95 |

*previously reported
†MMS 3 data outage

**Table 1a):** The thirty-two selected EDR or near-EDR encounters, sorted in chronological order, by row. These quantities were first computed at the individual spacecraft level, over a 4-second interval spanning +/- 2 s around each event's "central time" (see Sect. 4). Each spacecraft's result was then averaged to produce the numbers provided here. Column 1 (C1) lists the event name assignments, and Column 2 (C2) designates each event's central time. C3, C4, & C5 are the X, Y, and Z positions of the MMS centroid, in GSM coordinates, and C6 is the spacecraft separation. The maximum electron gyroradius is listed in C7, and C8 gives the max for the ion gyroradius. C9 is the average electron number density. C10 & C11 are the average and maximum current densities, respectively, C12 is the maximum value of the GSM y-component of the current, and C13 is the smallest B-field magnitude.

b)

| Event | Date & Time (UTC) | Max $E_\parallel$ (mV*m⁻¹) | Avg. $T_e$ (eV) | Min $T_{e\parallel}/T_{e\perp}$ (ratio) | Max $T_{e\parallel}/T_{e\perp}$ (ratio) | Avg. j*E' (nW*m⁻³) | Max j*E' (nW*m⁻³) | Int. j*E' (nW*m⁻³) | Avg. √$Q_e$ (index) | Max √$Q_e$ (index) |
|---|---|---|---|---|---|---|---|---|---|---|
| A01* | 2015/09/19 07:43:30 | 16.45 | 46.4 | 0.87 | 2.01 | 0.057 | 6.05 | 12.70 | 0.017 | 0.060 |
| A02* | 2015/10/16 10:33:30 | 7.80 | 36.6 | 0.84 | 2.00 | 0.244 | 2.42 | 6.66 | 0.017 | 0.052 |
| A03* | 2015/10/16 13:07:02 | 105.94 | 66.1 | 0.92 | 2.22 | 0.117 | 22.57 | 26.24 | 0.030 | 0.090 |
| A04* | 2015/10/22 06:05:22 | 48.06 | 51.8 | 0.66 | 1.83 | -0.342 | 7.47 | 16.93 | 0.018 | 0.069 |
| A05* | 2015/11/01 15:08:06 | 29.44 | 53.2 | 0.68 | 2.01 | 0.217 | 4.15 | 8.95 | 0.023 | 0.072 |
| A06* | 2015/11/12 07:19:21 | 2.82 | 45.8 | 0.90 | 1.75 | 0.002 | 0.97 | 0 | 0.018 | 0.086 |
| A07* | 2015/12/06 23:38:31 | 109.77 | 111.9 | 0.58 | 2.81 | 0.563 | 10.13 | 23.88 | 0.022 | 0.066 |
| A08* | 2015/12/08 11:20:44 | 60.68 | 86.1 | 0.97 | 4.43 | 0.163 | 8.31 | 18.01 | 0.026 | 0.084 |
| A09 | 2015/12/09 01:06:11 | 20.91 | 64.2 | 0.95 | 2.46 | -0.422 | 1.07 | 0 | 0.019 | 0.051 |
| A10* | 2015/12/14 01:17:40 | 63.73 | 103.1 | 0.93 | 2.80 | 0.577 | 7.13 | 15.16 | 0.031 | 0.095 |
| A11 | 2016/01/07 09:36:15 | 3.30 | 67.0 | 1.05 | 1.81 | 0.759 | 6.78 | 32.95 | 0.017 | 0.047 |
| A12* | 2016/01/10 09:13:37 | 51.93 | 73.1 | 0.69 | 2.39 | 0.924 | 13.98 | 55.33 | 0.022 | 0.066 |
| A13* | 2016/02/07 20:23:35 | 7.66 | 56.5 | 0.79 | 1.48 | 0.055 | 0.38 | 0 | 0.016 | 0.057 |
| B14 | 2016/10/22 12:58:41 | 52.27 | 27.1 | 0.90 | 2.80 | 0.848 | 11.92 | 61.31 | 0.019 | 0.055 |
| B15 | 2016/11/02 14:46:18 | 16.38 | 46.9 | 0.89 | 1.65 | -0.089 | 1.09 | 0 | 0.016 | 0.036 |
| B16 | 2016/11/06 08:40:58 | 44.14 | 58.7 | 1.54 | 2.97 | 0.445 | 8.12 | 20.52 | 0.037 | 0.075 |
| B17 | 2016/11/12 17:48:47 | 38.55 | 133.5 | 0.77 | 1.97 | -0.213 | 5.12 | 6.24 | 0.014 | 0.048 |
| B18 | 2016/11/13 09:10:41 | 36.17 | 88.9 | 0.81 | 1.51 | 0.206 | 18.28 | 48.20 | 0.011 | 0.050 |
| B19 | 2016/11/18 12:08:11 | 5.10 | 42.7 | 0.99 | 2.93 | 0.043 | 1.02 | 0 | 0.016 | 0.050 |
| B20 | 2016/11/23 07:49:33 | 8.83 | 67.8 | 0.71 | 3.49 | 0.445 | 7.35 | 19.47 | 0.024 | 0.121 |
| B21 | 2016/11/23 07:49:52 | 133.36 | 94.5 | 1.19 | 3.78 | 3.130 | 32.32 | 124.23 | 0.031 | 0.089 |
| B22 | 2016/11/23 07:50:30 | 33.59 | 86.6 | 0.88 | 3.55 | 0.733 | 8.57 | 26.44 | 0.031 | 0.078 |
| B23 | 2016/11/28 15:47:00 | 11.82 | 45.0 | 1.04 | 3.10 | 0.077 | 1.63 | 0.63 | 0.015 | 0.036 |
| B24 | 2016/12/11 04:41:50 | 2.85 | 77.7 | 0.73 | 1.99 | 0.342 | 2.07 | 2.20 | 0.010 | 0.026 |
| B25 | 2016/12/19 14:15:02 | 27.59 | 61.6 | 0.95 | 3.06 | 0.146 | 2.33 | 2.33 | 0.016 | 0.050 |
| B26† | 2017/01/02 02:58:13 | 8.24 | 57.2 | 0.65 | 1.55 | 0.375 | 2.37 | 21.98 | 0.012 | 0.026 |
| B27 | 2017/01/11 04:22:43 | 5.07 | 67.1 | 0.85 | 1.33 | 0.416 | 8.53 | 25.98 | 0.010 | 0.055 |
| B28 | 2017/01/20 12:32:07 | 97.06 | 90.2 | 0.97 | 2.62 | 1.480 | 8.31 | 78.36 | 0.015 | 0.051 |
| B29 | 2017/01/22 10:15:46 | 19.70 | 70.4 | 0.58 | 2.36 | 0.414 | 4.13 | 20.23 | 0.018 | 0.043 |
| B30 | 2017/01/22 10:15:58 | 10.19 | 72.7 | 0.94 | 2.26 | 0.214 | 3.39 | 5.40 | 0.017 | 0.044 |
| B31 | 2017/01/22 10:47:33 | 16.16 | 92.1 | 0.82 | 1.43 | 0.179 | 1.98 | 1.38 | 0.011 | 0.036 |
| B32 | 2017/01/27 12:05:43 | 12.65 | 95.1 | 0.90 | 1.61 | 0.447 | 7.51 | 62.75 | 0.015 | 0.046 |

*previously reported
†MMS 3 data outage

**Table 1b):** A continuation of Table 1a. Column 1 (C1) re-lists the event name assignments, and Column 2 (C2), the "central" times. C14 is the maximum E-field measured parallel to **B**, and C15 is the average electron temperature. C16 & C17 are the minimum and maximum ratios of electron temperature parallel and perpendicular to **B**. C18 is the average **j** * **E**', C19 the max, and C20 is the integrated **j** * **E**', defined as the accrued **j** * **E**' over the longest consecutive set of 30 ms intervals exceeding a threshold of 2 nW*m⁻³. Some events do not contain **j** * **E**' > 2 observed by any spacecraft, translating into "0" values. C21 and C22 are the average and maximum agyrotropy. Agyrotropy measurements for data points of $n_e$ < 5 cm⁻³ were not included in our computations (Sect. 4). We note again: all quantities are the averages across the four spacecraft.

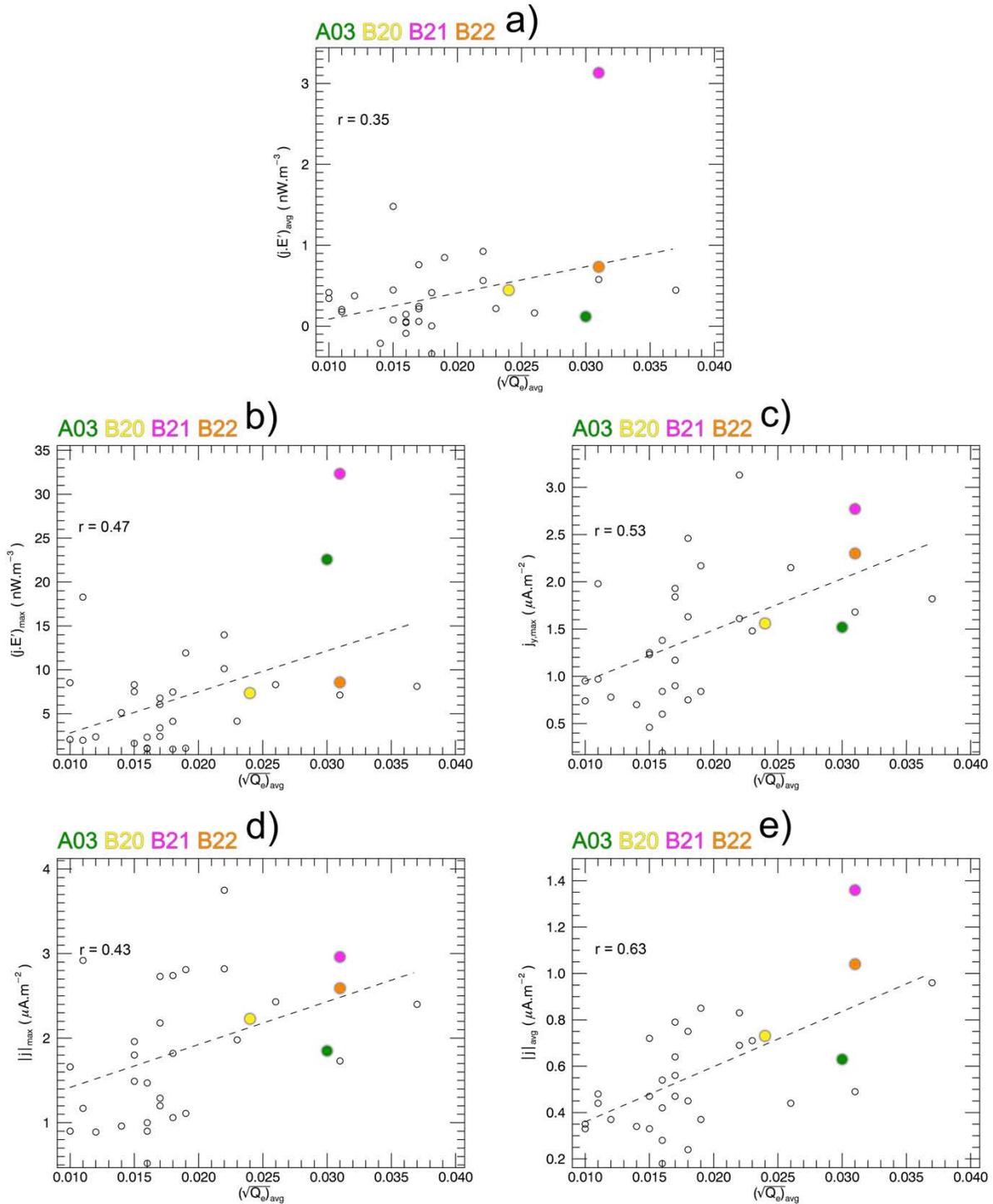

**Figure 8:** Five quantities correlated against the 4 s average of $\sqrt{Q_e}$. All values are taken from Table 1. **a)** shows the avg. of **j** * **E**' as a function of $\sqrt{Q_e}$, **b)** is max **j** * **E**' vs. avg. $\sqrt{Q_e}$, **c)** is max +$j_y$ vs. avg. $\sqrt{Q_e}$, **d)** is max |**j**| vs. avg. $\sqrt{Q_e}$, and **e)** is avg. |**j**| vs. avg. $\sqrt{Q_e}$. Four data points in each plot are colored, corresponding to Event A03 (= green), Event B20 (= yellow, previously Event "1"), Event B21 (= magenta, previously Event "2"), and Event B22 (= orange, previously Event "3").